\documentclass[11pt, a4paper]{article}
\usepackage[textwidth = 7in, textheight=25cm,nohead]{geometry}
\usepackage{ulem}
\usepackage{graphicx, amssymb, amsmath, amsthm, amstext, booktabs, float, multirow, subfigure, rotating,natbib}
\usepackage{enumitem}

\usepackage{setspace}
\numberwithin{equation}{section} 
\numberwithin{figure}{section} 
\numberwithin{table}{section}

\theoremstyle{remark}

\bibpunct{(}{)}{,}{a}{,}{,}

\newcommand{{\toL}}{{\ \overset{\mathcal{L}}{\longrightarrow}\ }}

\def\ppn{\vskip 6pt \noindent }
\def\R{{\mathbb{R}}}

\def\P{{\mathbb{P}}}
\def\E{{\mathbb{E}}}
\newcommand{{\Xs}}{{\cal X}}
\newcommand{{\Ls}}{{\cal L}}
\newcommand{{\Ss}}{{\cal S}}
\newcommand{{\Us}}{{\cal U}}
\newcommand{{\Gs}}{{\cal G}}
\newcommand{{\Hs}}{{\cal H}}
\newcommand{{\Ns}}{{\cal N}}
\newcommand{{\Is}}{{\cal I}}
\newcommand{{\Bs}}{{\cal B}}
\newcommand{{\Cs}}{{\cal C}}
\newcommand{{\Rs}}{{\cal R}}
\newcommand{{\pp}}{{\mathbf p}}
\newcommand{{\KK}}{{\mathbf K}}
\newcommand{{\HH}}{{\mathbf H}}
\newcommand{{\II}}{{\mathbf I}}
\newcommand{{\yy}}{{\mathbf y}}
\newcommand{{\YY}}{{\mathbf Y}}
\newcommand{{\dou}}{$\leadsto$\ }

\DeclareMathOperator{\logit}{logit}

\begin{document}

\title{Robust analysis of second-leg home advantage in UEFA football through better nonparametric confidence intervals for binary regression functions}
\author{\sc{Gery Geenens}\thanks{Corresponding author: ggeenens@unsw.edu.au, School of Mathematics and Statistics, UNSW Sydney, Australia, tel +61 2 938 57032, fax +61 2 9385 7123 }\\School of Mathematics and Statistics, \\ UNSW Sydney, Australia \and  \sc{Thomas Cuddihy} \\ School of Mathematics and Statistics,\\ UNSW Sydney, Australia}
\date{\today}
\maketitle
\thispagestyle{empty} 

\begin{abstract} 
In international football ({\it soccer}), two-legged knockout ties, with each team playing at home in one leg and the final outcome decided on aggregate, are common. Many players, managers and followers seem to believe in the `second-leg home advantage', i.e.\ that it is beneficial to play at home on the second leg. A more complex effect than the usual and well-established home advantage, it is harder to identify, and previous statistical studies did not prove conclusive about its actuality. Yet, given the amount of money handled in international football competitions nowadays, the question of existence or otherwise of this effect is of real import. As opposed to previous research, this paper addresses it from a purely nonparametric perspective and brings a very objective answer, not based on any particular model specification which could orientate the analysis in one or the other direction. Along the way, the paper reviews the well-known shortcomings of the Wald confidence interval for a proportion, suggests new nonparametric confidence intervals for conditional probability functions, revisits the problem of the bias when building confidence intervals in nonparametric regression, and provides a novel bootstrap-based solution to it. Finally, the new intervals are used in a careful analysis of game outcome data for the UEFA Champions and Europa leagues from 2009/10
to 2014/15. A slight `second-leg home advantage' is evidenced.
\end{abstract}

\section{Introduction}\label{sec:intro}

The `{\it home field advantage}' in sport is well established, as its quantitative study can be traced back to the late 70's \citep{Schwartz1977}. A meta-analysis of 30 research articles by \cite{Jamieson2010}, including over 30,000 games, found significant home field advantages in each of 10 different sports, be it individual such as tennis or team such as football ({\it soccer}). Yet, there seems to exist a downward trend in this advantage over time. \cite{Pollard2005a} analysed more than 400,000 games, dating back to 1876, from ice hockey, baseball, American football, basketball and football, and found a decline in home field advantage for most sports. Nevertheless, the advantage was still positive for all sports in the final year analysed (2002). 

\ppn The exact causes of that advantage are multiple and complex. In an effort to better understand them, an important literature review by \cite{Courneya1992} developed a conceptual framework, which was later updated by \cite{Carron2005}. It incorporates 5 major components: game location; game location factors; critical psychological and physiological states; critical behavioural states; and performance outcomes. \cite{Carron2005} posited that each component affects all subsequent ones. For example, `game location' influences `game location factors' such as crowd size and composition, which in turn affects the `psychological state' and then `behavioural state' of the players, etc. They concluded that further research is necessary in all of the 5 components to better understand the individual impacts, for example there is insufficient research into the effect of crowd density and absolute crowd size and their interaction. More recently, \cite{Pollard2008} reached a similar conclusion, and reiterated their thoughts from an analysis two decades earlier \citep{Pollard1986}: ``{\it Clearly, there is still much to be learnt about the complex mechanisms that cause home advantage, both in soccer and other sports. The topic remains a fruitful area of research for sports historians, sociologists, psychologists and statisticians alike.}''

\ppn One aspect of the home field advantage whose existence is still being debated is that of the `{\it second-leg home advantage}' (hereafter: SLHA), when a contest between two teams comprises two matches (`legs'), with each team as the home team in one leg, and the final outcome decided on aggregate. At first sight, one might expect that each team would get their respective home advantage and that the effects would cancel out. Yet, a common belief is that the team playing at home for the second leg has a slight advantage over the other team. One theory to support this claim is that the team playing away on the first game can essentially play it safe there, while taking advantage of playing at home on the decider game when the difference is to be made. The stake of the second leg being higher, the crowd support and the induced pressure might indeed be more intense then than on the first leg, where getting the upper hand is unlikely to be final anyway. This may create an asymmetry in home advantage between the two legs \citep{Pollard2006a,Page2007a}.

\ppn Those two-legged ties are very common in knockout stages of football international club competitions, such as national team play-offs in some qualification tournaments, including the FIFA World Cup, and most prominently the European cups, namely the UEFA Champions League and Europa League. Those competitions are big business, especially the UEFA Champions League. For instance, from the season 2015-2016 onwards, to qualify to the quarter-finals brings in a bonus of 6 millions of euros; to advance further to the semi-finals brings in an extra bonus of 7 millions of euros; and to qualify to the final brings in another 10.5 or 15 millions of euros (depending on the outcome of the final)\footnote{{\tt http://www.uefa.com/uefachampionsleague/news/newsid=2398575.html} -- these amounts are the `fixed amounts', awarded according to the clubs' performance; they excluded the so-called `market pool', which essentially comes from television income.}. Hence, beyond the sporting aspect, the difference between qualifying or being eliminated at some level of the knockout stage represents a huge amount of money. As a result, an unwarranted advantage for the team playing at home second implies an equally unwarranted economic shortfall for the team playing at home first, whose only fault is to have been unlucky at the draw. The question of existence of the SLHA is, therefore, of great significance.

\ppn Consequently, scholars have attempted to evidence or otherwise the SLHA, with research focussing on the case of UEFA administered competitions. Yet, none really proved conclusive. A naive comparison of the fraction of teams qualifying when playing at home on the second leg against when playing at home on the first leg, is not telling. This is because of the non-random manner in which UEFA sometimes seeds teams in knockout stages of their competitions. For instance, in the first knockout round following the initial group stage, such as the Round-of-16 in the Champions league or Round-of-32 in the Europa league, group winners (supposedly the best teams) are automatically assigned to play the second game at home. Obviously, this may lead to the spurious result of existence of SLHA, as stronger teams are preferentially allocated as playing at home second in some instances. This induces an obvious confounding factor.

\ppn \cite{Page2007a} and \cite{Eugster2011} adjusted for it by conditioning their analyses on the difference between the UEFA coefficients of the two teams, assumed to be a reasonable proxy for their relative strength (more on this in Section \ref{sec:chap4}). \cite{Eugster2011} found no evidence of SLHA and \cite{Page2007a} found a significant effect in seasons before 1994/95 but not afterwards. By contrast, \cite{Lidor2011} and \cite{Flores2015} did find a significant SLHA effect, however they did not control for the confounding factor, which greatly lowers the value of their study. It seems indeed clear that the relative strength of the matched teams should be taken into account for meaningful analyses. Hence this paper will focus on the `conditional' case only.

\ppn Specifically, denote $p(x)$ the probability of the second-leg home team qualifying given that the difference in `strength' between the two teams (however it is measured) at the time of the game is $x$. Then, $p(0)$ is the probability of the second-leg home team going through, given that the two teams are of the same strength. The probability $p(0)$ is, therefore, the parameter of main interest in this study, with a value for $p(0)$ above $1/2$ indicating a second-leg home advantage. Existence of the SLHA effect can thus be formally tested by checking whether an empirical estimate $\hat{p}(0)$ of $p(0)$ significantly lies above $1/2$.

\ppn \cite{Page2007a} and \cite{Eugster2011} estimated $p(x)$ using a logistic regression model, but failed to provide an examination of goodness-of-fit of this parametric model and just stated regression estimates as-is. To offer an alternate viewpoint on the question, this paper explores the topic using nonparametric techniques, really `letting the data speak for themselves'. To this effect, a Nadaraya-Watson (NW) kernel regression model \citep[Section 4.1.2]{Hardle2004} has been developed to regress the final outcome of the two-legged knockout on a measure of the inequality in team strengths, based again on the UEFA club coefficients. This model allows robust and objective estimation of $p(x)$ from historical data. A measure of statistical significance is provided by the asymptotic normality of the Nadaraya-Watson estimator, which enables the calculation of pointwise confidence intervals for $p(x)$ at any $x$ \citep[Section 4.4.1]{Hardle2004}. 

\ppn However, the working model being essentially a conditional Bernoulli model here, the estimated probability $\hat{p}(x)$ is some kind of `conditional sample proportion' (see Section \ref{sec:chap3} for details). The above standard confidence intervals for $p(x)$ thus amount to some sort of Wald intervals, adapted to the conditional case. In the classical (i.e., non-conditional) framework, it is well known  \citep{Agresti1998} that the coverage of the Wald interval for a proportion $p$ can be very poor, and this for any values of $p \in (0,1)$ and the sample size $n$. \cite{Brown2001b,Brown2002} explained how the coverage probability of the Wald interval is affected by both systematic negative bias and oscillations, 
and recommended three alternatives to the Wald interval: the Wilson, the Agresti-Coull and the Jeffreys intervals. 

\ppn The methodological contribution of this paper is twofold. First, such `better' confidence intervals will be obtained for a conditional probability estimated by the Nadaraya-Watson estimator. `Conditional' versions of the Wilson and Agresti-Coull confidence intervals will thus be constructed. When doing so, the inherent bias of the NW estimator will be a major factor to take into account, as often in procedures based on nonparametric function estimation. Consequently, the second contribution will be to devise a careful strategy for efficiently dealing with that bias when computing the above confidence intervals in practice. Finally, those will be used for the interval-estimation of the probability $p(0)$ of interest. This will allow the research question about the existence of some SLHA to be addressed in a very robust way. 

\ppn The paper is organised as follows. Section \ref{sec:chap2} provides an overview of the work by \cite{Agresti1998} and \cite{Brown2001b,Brown2002} about confidence intervals for a proportion. A particular emphasis will be on the Wilson and Agresti-Coull confidence intervals, in order to prepare for their adaptation to the conditional case in Section \ref{sec:chap3}. 
There, some background on the standard, Wald-type interval for a conditional probability estimated by the Nadaraya-Watson estimator is provided. Then details of the derivations of the Wilson and Agresti-Coull intervals for the conditional case are given. The problem of the bias is addressed, and a novel way of choosing the right smoothing parameter when constructing the confidence intervals is suggested. The performance of the new intervals based on this strategy are then analysed through a simulation study. Section \ref{sec:chap4} comes back to the research question and presents the results. A discussion about the implications and limitations of the current study follows, with some suggestions for future research in the field. Finally, Section \ref{sec:chap5} concludes.

\section{Confidence intervals for a proportion} \label{sec:chap2}

\subsection{Background} \label{subsec:WaldCI}

Consider a random sample $\{Y_1,Y_2,\ldots,Y_n\} \overset{\text{i.i.d.}}{\sim}$ Bernoulli$(p)$, for some value $p \in (0,1)$ (degenerate cases $p=0$ or $p=1$ have very limited interest), and say that $Y_i = 1$ if individual $i$ has a certain characteristic, and $Y_i=0$ otherwise. Denote $\YY \doteq \sum_{i=1}^n Y_i$, the number of sampled individuals with the characteristic. The sample proportion $\hat{p} = \frac{\sum_{i=1}^n Y_i}{n} = \frac{\YY}{n}$ is known to be the maximum likelihood estimator of $p$, satisfying
\begin{equation} \sqrt{n}(\hat{p} - p) \toL \mathcal{N}(0,p(1-p)) \label{eqn:asymptnorm}\end{equation} 
as $n \to \infty$ from the Central Limit Theorem. Substituting in $\hat{p}$ for the estimation of the standard error, a confidence interval of level $1-\alpha$ for $p$ easily follows:
\begin{equation} \overline{CI}_\text{Wa} = \left[\hat{p} \pm z_{1-\alpha/2} \sqrt{\frac{\hat{p}(1-\hat{p})}{n}}\,\right], \label{eqn:WaldCI} \end{equation}
where $z_{a}$ is the quantile of level $a \in (0,1)$ of the standard normal distribution. This is the {\it Wald interval}. 

\ppn Unfortunately, this interval reputedly shows poor behaviour in terms of coverage probability, as has been known for quite some time, see for instance \cite{Cressie1978,Ghosh1979} and \cite{Blyth1983}. \cite{Brown2001b,Brown2002} provided a thorough understanding of the cause of the phenomenon. They showed, even for large values of $n$, that the pivotal function $W=\frac{n^{1/2}(\hat{p} - p)}{\sqrt{\hat{p}(1-\hat{p})}}$ can be significantly non-normal with large deviations of bias, variance, skewness and kurtosis. Of particular importance is the bias term \citep[Section 2.1]{Brown2002}:
\begin{equation}
\E(W)=\E\left(\dfrac{n^{1/2}(\hat{p} - p)}{\sqrt{\hat{p}(1-\hat{p})}}\right)= \frac{p - 1/2}{\sqrt{np(1-p)}}\left(1 + \frac{7}{2n}+\frac{9(p-1/2)^2}{2np(1-p)}\right) + o(n^{-3/2}), \label{eqn:biasW}
\end{equation}
which is non-zero for $p \neq \frac{1}{2}$ and changes sign depending if $p<\frac{1}{2}$ or $p>\frac{1}{2}$. Hence, even though $\hat{p}$ is unbiased for $p$, the estimation of the standard error by substituting in $\hat{p}$ introduces some substantial positive or negative bias in $W$. This eventually results in systematic negative bias for the coverage probability of the interval based on $W$, a correction of which requiring a shift of the centre of the interval towards $\frac{1}{2}$. Obviously, this problem is most serious for values of $p$ `far away' from $1/2$, i.e.\ close to 0 or 1. This is the origin of the popular rule-of-thumb `$np(1-p)$ must be greater than 5 (or 10)', supposed to validate the usage of the Wald interval, but mostly discredited by (\ref{eqn:biasW}). In addition, the coverage probability also suffers from important oscillations across $n$ and $p$ \citep[Figure 1]{Brown2002}. This is essentially due to the discrete nature of $\YY$: clearly, for finite $n$, $\hat{p}$ can only take on a finite number of different values ($0$, $1/n$, $2/n$, $\ldots$, $(n-1)/n$, $1$), and that causes problems when approximating its distribution by a normal smooth curve, see \citet[Section 2.2]{Brown2002} for details.

\ppn Hence \cite{Brown2002} went on to investigate a dozen of alternatives, including the `exact' Clopper-Pearson interval, an interval based on the Likelihood Ratio test, and some intervals based on transformations such as arcsine and logit. They recommended only three of them as replacements for the Wald interval: the Wilson, the Agresti-Coull and the Jeffreys intervals. This is due to their superior performance in coverage probability and interval length, as well as their ease of interpretation. 

\ppn In this paper the Wilson and Agresti-Coull intervals will be adapted to the conditional, kernel regression-based, setting. These two intervals are derived from the asymptotic normality of the sample proportion (\ref{eqn:asymptnorm}). As it is known, the Nadaraya-Watson estimator of the conditional probability $p(x)$, that will be used in Sections \ref{sec:chap3} and \ref{sec:chap4}, is a weighted sample average, which can be regarded as a `conditional sample proportion' in this framework. In particular, the Nadaraya-Watson estimator is - under mild conditions - asymptotically normally distributed as well \citep[Theorem 4.5]{Hardle2004}, which allows a natural extension of the Wilson and Agresti-Coull intervals to the conditional case. On the other hand, the Jeffreys interval has a Bayesian derivation, being essentially a credible interval from the posterior distribution of $p$ when some uninformative (`{\it Jeffreys}') prior is used. It is less obvious how this construction would fit in the conditional setting. It seems, therefore, reasonable to leave the Jeffreys interval on the side in this paper. 

\subsection{The Wilson and Agresti-Coull confidence intervals for a proportion} \label{subsec:WiAC}

The Wilson interval, first described by \cite{Wilson1927}, follows from the inversion of the score test for a null hypothesis $H_0: p=p_0$, hence it is also known as `score interval'.  Like the Wald interval, it is obtained from (\ref{eqn:asymptnorm}). The main difference, though, is that the variance factor $p(1-p)$ is not estimated and keeps its unknown nature through the derivation. Specifically, from a statement like
\[\P\left(-z_{1-\alpha/2} < \frac{\hat{p}-p}{\sqrt{\frac{p(1-p)}{n}}} <  z_{1-\alpha/2}\right) \simeq 1- \alpha, \]
essentially equivalent to (\ref{eqn:asymptnorm}), it follows that the confidence interval for $p$ at level $1-\alpha$ should be the set of all values of $p$ such that
\[(\hat{p}-p)^2 \leq\ \frac{p(1-p)}{n} z_{1-\alpha/2}^2.\]
Solving this quadratic inequality in $p$ yields
\begin{equation} \overline{CI}_{\text{Wi}} = \left[\dfrac{\hat{p} + \frac{z_{1-\alpha/2}^2}{2n} \pm \frac{z_{1-\alpha/2}}{n^{1/2}} \sqrt{\hat{p}(1-\hat{p}) + \frac{z_{1-\alpha/2}^2}{4n}}}{1+\frac{z_{1-\alpha/2}^2}{n}}\,\right] = \left[\frac{\YY + z_{1-\alpha/2}^2/2}{n+z_{1-\alpha/2}^2} \pm \frac{n^{1/2}z_{1-\alpha/2}}{n+z_{1-\alpha/2}^2} \sqrt{\hat{p}(1-\hat{p}) + \frac{z_{1-\alpha/2}^2}{4n}} \,\right]. \label{eq:wilson_interval_nonCond}\end{equation}
\cite{Olivier2006} showed that this interval can be written
\begin{equation} \overline{CI}_{\text{Wi}} = \left[(1-w)\hat{p} + (w) \frac{1}{2} \pm z_{1-\alpha/2} \sqrt{(1-w) \frac{\hat{p}(1-\hat{p})}{n+z^2_{1-\alpha/2}} + (w)\frac{1}{4(n+z^2_{1-\alpha/2})}}\,\right], \label{eqn:CIWiOM} \end{equation}
where $w=\frac{z_{1-\alpha}^2}{n+z_{1-\alpha}^2}$. This interval is symmetric around $(1-w)\hat{p} + (w) \frac{1}{2}$, a weighed average of $\hat{p}$ and the uninformative prior $\frac{1}{2}$, with the weight on $\hat{p}$ heading to 1 asymptotically. Compared to the Wald interval (\ref{eqn:WaldCI}), the interval centre is now shifted towards 1/2 which substantially reduces the bias in coverage probability as suggested below (\ref{eqn:biasW}). Likewise, the coefficient on $z_{1-\alpha/2}$ in the $\pm$ term is the same weighted average of the variance when $p=\hat{p}$ and when $p = \frac{1}{2}$. \citet[p.\,122]{Agresti1998} `strongly recommend' the Wilson interval as alternative to the Wald interval. However, they acknowledged that the form of this interval is complicated and so suggested a new interval with a simple form which they called the `Adjusted Wald interval', now better known as the `Agresti-Coull' interval.

\ppn They noted that the Wilson interval is like a `shrinking' of both the midpoint and variance estimate in the Wald interval towards $\frac{1}{2}$ and $\frac{1}{4}$ respectively, with the amount of shrinkage decreasing along with  $n$. This lead them to consider the midpoint of the Wilson interval, $\tilde{p} \doteq \frac{\YY + z_{1-\alpha/2}^2/2}{n + z_{1-\alpha/2}^2}$, as another point estimate of $p$, and then continue with the Wald Interval derivation. For the `usual' confidence level $1-\alpha = 0.95$, $z_{1-\alpha/2} =1.96 \simeq 2$, hence $\tilde{p} \simeq \frac{\YY + 2}{n + 4}$ and this procedure is sometimes loosely called the `{\it add 2 successes and 2 failures}' strategy. It combines the idea of shifting the centre toward $1/2$, {\it \`{a} la} Wilson, with the simplicity of the Wald interval derivation which substitutes in an estimate for the standard error.

\ppn Specifically, define 
\begin{equation} \widetilde{\YY} \doteq \YY + z_{1-\alpha/2}^2/2, \quad  \tilde{n} \doteq  n + z_{1-\alpha/2}^2 \quad \text{ and } \quad \tilde{p} \doteq \frac{\widetilde{\YY}}{\tilde{n}}. \label{eqn:ptilde} \end{equation}
Then, given that $\tilde{p}-\hat{p} = O(n^{-1})$, it follows from (\ref{eqn:asymptnorm}) that
\[\sqrt{n}\left( \tilde{p}-p\right) \toL \mathcal{N} \left(0, \frac{p(1-p)}{n}\,\right), \]
and acting as in Section \ref{subsec:WaldCI} yields 
\begin{equation} \overline{CI}_\text{AC} =\left[ \tilde{p}  \pm z_{1-\alpha/2}\sqrt{\frac{\tilde{p}(1-\tilde{p})}{\tilde{n}}}\,\right]. \label{eq:agresti_interval_nonCond_alt} \end{equation}

\ppn \cite{Brown2001b} provided an excellent breakdown of the performance of the Wald, Wilson and Agresti-Coull intervals as a whole. The Wilson and Agresti-Coull intervals behave very similarly. In particular, they are almost indistinguishable around $p=0.5$ \citep[Figure 5]{Brown2001b}. Most importantly, the Wilson and Agresti-Coull intervals maintain a coverage probability very close to the confidence level $1-\alpha$, and this for all $p$ and even for small values of $n$, as opposed to the Wald interval whose coverage probability remains consistently below target, even for large $n$. The oscillations persist, though, which could be expected: it is the discreteness of the Bernoulli distribution that causes the oscillations, not how the intervals are formed. \citet[Theorem 7]{Brown2002} also proved that both the Agresti-Coull and the Wilson intervals are shorter on average than the Wald interval. In the following section, conditional versions of these intervals are constructed, and it is analysed if these observations carry over to the case of estimating a conditional probability via nonparametric binary regression.

\section{Confidence intervals for a conditional probability} \label{sec:chap3}

\subsection{Binary regression and Nadaraya-Watson estimator}

Consider now a bivariate sample $\Xs=\{(X_1,Y_1),\ldots,(X_n,Y_n)\}$ of i.i.d.\ replications of a random vector $(X,Y) \in \R \times \{0,1\}$ such that $X \sim F$ (unspecified) and $Y |X \sim \text{Bernoulli}(p(X))$. Now, the probability of a certain individual having the characteristic of interest is allowed to vary along with another explanatory variable $X$. Of interest is the estimation of the conditional probability function
\begin{equation} p(x) = \P(Y=1 | X=x). \label{eqn:px} \end{equation}
Assuming that $X$ is a continuous variable whose distribution admits a density $f$, the estimation of $p(x)$ actually falls within the topic of regression. Indeed, 
\[ \E(Y|X=x) = 0\times(1-p(x)) + 1\times p(x) = p(x), \]
meaning that $p(x)$ is actually a conditional expectation function. The problem is called {\it binary regression}.

\ppn Common parametric specifications for $p$ include logistic and probit models. Their use is so customary that their goodness-of-fit is often taken for granted in applied studies. E.g., within the application considered in this paper, \cite{Page2007a} and \cite{Eugster2011} did not attempt any validation of their logistic model. The primary tool for suggesting a reasonable parametric specification in the `continuous-response' context is often the basic $(X,Y)$-graph (scatter-plot). When the response is binary, though, a scatter-plot is not much informative (no clear shape for the cloud of data points, see for instance Figure \ref{fig:phat} below), hence binary regression actually lacks that convenient visual tool. Maybe that is the reason why the question of goodness-of-fit of a logistic regression model is so often overlooked in the literature, as if the incapacity of visually detecting departures from a model automatically validates it. Yet, without any visual guide, the risk of model misspecification is actually higher in binary regression than in other cases \citep{Horowitz2001}, with misspecification typically leading to non-consistent estimates, biased analyses and questionable conclusions. 

\ppn In order to avoid any difficulty in postulating and validating some parametric specification for the function $p$, here a Nadaraya-Watson kernel regression estimator will be used. Kernel smoothing is a very popular {\it nonparametric} regression method \citep[Chapter 4]{Hardle2004}, and the Nadaraya-Watson (NW) estimator \citep{Nadaraya1964,Watson1964} one of its simplest variants. Given the sample $\Xs$, the NW estimator is defined as
\begin{equation} \hat{p}_h(x) = \frac{\sum_{i=1}^n K\left(\frac{x-X_i}{h} \right)Y_i}{\sum_{i=1}^n K\left(\frac{x-X_i}{h} \right)}, \label{eqn:NWest} \end{equation}
where $K$ is a `kernel' function, typically a smooth symmetric probability density like the standard Gaussian, and $h$ is a `bandwidth', essentially fixing the smoothness of the final estimate $\hat{p}_h$. Clearly, (\ref{eqn:NWest}) is just a weighted average of the binary values $Y_i$'s, with weights decreasing with the distance between $x$ and the corresponding $X_i$. Hence it returns an estimation of the `local' proportion of the $Y_i$'s equal to 1, for those individuals such that $X_i \simeq x$ \citep{Copas83}, which is indeed a natural estimate of (\ref{eqn:px}). It is straightforward to see that $\hat{p}_h(x)$ always belongs to $[0,1]$, as it is just a (weighted) average of 0/1 values. There exist more elaborated nonparametric regression estimators (Local Polynomial or Splines, for instance), but those usually fail to automatically satisfy this basic constraint on $p(x)$. Hence the NW estimator seems a natural choice here.

\ppn Classical results in kernel regression \citep[Theorem 4.1]{Hardle2004} state that estimator (\ref{eqn:NWest}) is a consistent one for $p(x)$ provided that $h \to 0$ and $nh \to \infty$ as $n \to \infty$. Moreover, if $h = O(n^{-1/5})$, it is asymptotically normal. Specifically, adapting Theorem 4.5 of \cite{Hardle2004} to the binary case gives, at all $x$ such that $f$ and $p$ are twice continuously differentiable, and $f(x) > 0$,
\begin{equation} \sqrt{nh}(\hat{p}_h(x) - p(x)) \toL \mathcal{N}\left(\frac{1}{2} \lambda \mu_2(K) b(x), R(K) \dfrac{p(x)(1-p(x))}{f(x)}\right), \label{eqn:asnormNW} \end{equation}
where $ \lambda  = \lim_{n\to\infty} nh^5  <\infty$, $ \mu_2(K) = \int u^2 K(u)\,du$ and $R(K) = \int K^2(u)\,du$ are kernel-dependent constants, and
\begin{equation} b(x) = p''(x) + \frac{2p'(x)f'(x)}{f(x)}. \label{eqn:bx} \end{equation}
Balancing squared bias and variance, in order to achieve minimum Mean Squared Error for the estimator, requires to take $h \sim n^{-1/5}$ \citep[Theorem 4.3]{Hardle2004}. This means $\lambda >0$, materialising a non-vanishing bias term  in (\ref{eqn:asnormNW}).

\ppn This bias has a major impact on any statistical procedure based on nonparametric function estimation \citep{Hall13}, and requires careful treatment. In the context of building confidence intervals, it can be either explicitly estimated and corrected \citep{Hardle1988,Eubank1993,Xia98}, or one can act via {\it undersmoothing}: if $h=o(n^{-1/5})$, hence purposely sub-optimal, then $\lambda = 0$ in (\ref{eqn:asnormNW}) which becomes 
\begin{equation} \sqrt{nh}(\hat{p}_h(x) - p(x))\toL \mathcal{N}\left(0, R(K) \dfrac{p(x)(1-p(x))}{f(x)}\right). \label{eqn:NW_asymp_normal_h_undersmooth} \end{equation}
The bias is seemingly gone; of course, at the price of an increased variance. \cite{Hall1992} and \cite{Neumann1997} theoretically demonstrated the superiority of treating the bias via undersmoothing over explicit correction in terms of empirical coverage of the resulting confidence intervals. Clearly, (\ref{eqn:NW_asymp_normal_h_undersmooth}) is the analogue of (\ref{eqn:asymptnorm}) for a conditional probability. It will consequently serve below as the basis for constructing confidence intervals for $p(x)$ at any $x$. 

\subsection{Wald interval for a conditional probability} \label{subsec:WaldCondCi}

In particular, a Wald-type confidence interval at level $1-\alpha$ for $p(x)$ is
\begin{equation} 
CI_{\text{Wa}}(x;h)= \left[\hat{p}_h(x) \pm z_{1-\alpha/2} \sqrt{ \frac{\hat{p}_h(x)(1-\hat{p}_h(x))}{nh\hat{f}_h(x)/R(K)}}\,\right]. \label{eq:wald_kernel}
\end{equation}
It directly follows from the asymptotic normality statement (\ref{eqn:NW_asymp_normal_h_undersmooth}) with the variance $R(K) \frac{p(x)(1-p(x))}{f(x)}$ being estimated: $p(x)$ is estimated by $\hat{p}_h(x)$ (\ref{eqn:NWest}), and $f(x)$ is estimated by its classical kernel density estimator \citep[Chapter 3]{Hardle2004}
\begin{equation}
\hat{f}_{h}(x) = \frac{1}{nh}\displaystyle\sum_{i=1}^{n}K\left(\frac{x-X_i}{h}\right). \label{eqn:fhat}
\end{equation}
Although it would not necessarily be optimal for estimating $f(x)$ itself, here the same kernel $K$ and bandwidth $h$ as in (\ref{eqn:NWest}) should be used in (\ref{eqn:fhat}). The reason why a factor $1/f(x)$ arises in the variance of (\ref{eqn:asnormNW})/(\ref{eqn:NW_asymp_normal_h_undersmooth}) is that not all $n$ observations, but only a certain fraction (asymptotically) proportional to $f(x)$ are effectively used by the essentially local estimator (\ref{eqn:NWest}) for estimating $p$ at $x$. So, the estimation of $f$ here should be driven by accurately quantifying that `local equivalent sample size' at $x$. The fact that the quantity $nh \hat{f}_h(x)/R(K)$ is actually that equivalent sample size follows by seeing that $nh\hat{f}_h(x)$ is the denominator of (\ref{eqn:NWest}), while $R(K)$ gives an appreciation of how large is the weight given to those observations `close' to $x$, overall.

\ppn The Wald nature of (\ref{eq:wald_kernel}) and the shortcomings of its analogue exposed in Section \ref{subsec:WaldCI}, though, motivate the adaptation of the Wilson and Agresti-Coull intervals to the conditional context.

\subsection{Wilson and Agresti-Coull intervals for a conditional probability}

The derivation of the `conditional' Wilson interval follows the same steps and justification as those presented in Section \ref{subsec:WiAC}. From (\ref{eqn:NW_asymp_normal_h_undersmooth}), it is seen that a confidence interval of level $1-\alpha$ for $p(x)$ should wrap up all those values of $p(x)$ such that
\[ -z_{1-\alpha/2} < \frac{\hat{p}_h(x) - p(x)}{\sqrt{\frac{R(K)p(x)(1-p(x))}{nhf(x)}}} < \ z_{1-\alpha/2}, \]
that is,
\[ (\hat{p}_h(x) - p(x))^2 \leq\ \frac{R(K)p(x)(1-p(x))}{nhf(x)} z_{1-\alpha/2}^2. \]
Solving for $p(x)$, and estimating the unknown $f(x)$ by its kernel estimator $\hat{f}_h$ (\ref{eqn:fhat}), yields the interval
\begin{align} CI_\text{Wi}(x;h) & = \left[ \frac{\hat{p}_h(x) + \frac{z_{1-\alpha/2}^2R(K)}{2nh\hat{f}_h(x)} \pm \frac{z_{1-\alpha/2} R(K)^{1/2}}{(nh\hat{f}_h(x))^{1/2}}  \sqrt{\hat{p}_h(x)(1-\hat{p}_h(x)) + \frac{z_{1-\alpha/2}^2R(K)}{4nh\hat{f}_h(x)}}}{1 + \frac{z_{1-\alpha/2}^2R(K)}{nh\hat{f}_h(x)} }\, \right] \notag \\
& = \left[ \frac{\frac{\hat{p}_h(x)nh\hat{f}_h(x)}{R(K)} + \frac{z_{1-\alpha/2}^2}{2}}{\frac{nh\hat{f}_h(x)}{R(K)}+z_{1-\alpha/2}^2}  \pm \frac{z_{1-\alpha/2} \left(\frac{nh\hat{f}_h(x)}{R(K)}\right)^{1/2}}{\frac{nh\hat{f}_h(x)}{R(K)}+z_{1-\alpha/2}^2} \sqrt{\hat{p}_h(x)(1-\hat{p}_h(x)) + \frac{z_{1-\alpha/2}^2R(K)}{4nh\hat{f}_h(x)}} \,\right]. \label{eqn:wilson_interval_alt} \end{align}
Similarly to (\ref{eqn:CIWiOM}), the centre of this interval can be represented as $(1-w)\hat{p}_h(x) + (w)\frac{1}{2}$, where $w = \frac{z_{1-\alpha/2}^2}{nh\hat{f}_h(x)/R(K)+z_{1-\alpha/2}^2}$. This highlights that the adaptation to the conditional case has not altered the interval's nature. In addition, \eqref{eqn:wilson_interval_alt} directly suggests an `Agresti-Coull', simpler version of it.

\ppn Indeed, the (non-conditional) Agresti-Coull interval \eqref{eq:agresti_interval_nonCond_alt} is built around $\tilde{p}$, the centre of the corresponding (non-conditional) Wilson interval (\ref{eq:wilson_interval_nonCond}). Extending this to the conditional case from (\ref{eqn:wilson_interval_alt}), one can define
\begin{equation} CI_\text{AC}(x;h) = \left[\widetilde{p}_h(x)  \pm z_{1-\alpha/2}\sqrt{\frac{\widetilde{p}_h(x)(1-\widetilde{p}_h(x))}{\tilde{n}_h(x)}}\,\right] \label{eqn:ACCIcond} \end{equation}
where
\[\widetilde{p}_h(x) = \frac{\frac{\hat{p}_h(x)nh\hat{f}_h(x)}{R(K)} + \frac{z_{1-\alpha/2}^2}{2}}{\frac{nh\hat{f}_h(x)}{R(K)}+z_{1-\alpha/2}^2} \qquad \text{ and } \qquad \tilde{n}_h(x) = \frac{nh\hat{f}_h(x)}{R(K)}+z_{1-\alpha/2}^2. \]
The interpretation of $\frac{nh\hat{f}_h(x)}{R(K)}$ as the `local equivalent sample size' makes the analogy between this and (\ref{eqn:ptilde}) obvious.

\subsection{Choice of `undersmoothed' bandwidth} \label{subsec:USh}

The three intervals (\ref{eq:wald_kernel}), (\ref{eqn:wilson_interval_alt}) and (\ref{eqn:ACCIcond}) follow straight from (\ref{eqn:NW_asymp_normal_h_undersmooth}), which holds true if and only if $h=o(n^{-1/5})$ ({\it undersmoothing}). Results of \cite{Chen2002} on asymptotic intervals in nonparametric regression, suggest that the coverage probability of the Wald-type interval (\ref{eq:wald_kernel}) is
\begin{equation} \P\left(p(x) \in CI_\text{Wa}(x;h)\right) = 1-\alpha + O\left(nh^5 + h^2 +(nh)^{-1}\right). \label{eqn:covprob}\end{equation}
Hence, for minimising the coverage error of the confidence interval computed on $\hat{p}_h(x)$, one should take $h$ such that $h \sim n^{-1/3}$. Common practice in nonparametric methods requiring such undersmoothing, is to take $h$ `smaller' than a value, say $h_0$, supposed to be optimal for estimating $p(x)$. Typically, $h_0$ would be returned by a data-driven selection procedure, such as cross-validation or plug-in, see \cite{Kohler14} for a review. Often, $h_0$ is then just divided by some constant which heuristically looks appropriate. In the present case, one could take $h = h_0 n^{-1/3}/n^{-1/5} = h_0 n^{-2/15}$, so as to (supposedly) obtain $h \sim n^{-1/3}$, given that $h_0 \sim n^{-1/5}$, see comments below (\ref{eqn:bx}).

\ppn There is actually no justification for doing so in practice. `Undersmoothing' is a purely asymptotic, hence theoretical, concept. The `undersmoothed' $h$ is to tend to 0 quicker than the optimal $h_0$ {\it as $n$ would tend to $\infty$}, but this convergence is obviously meaningless when facing a sample of data of fixed size $n$. Indeed expressions like $h \sim n^{-1/3}$ or $h_0 \sim n^{-1/5}$ do not really make sense for fixed $n$. It is understood that, mainly because of the inherent bias of $\hat{p}_h(x)$, the value of $h$ leading to confidence intervals with good coverage properties is not, in general, the optimal bandwidth $h_0$ for estimating $p(x)$. However asymptotic expressions such as (\ref{eqn:covprob}) cannot really be of any practical help. In fact, there are no effective empirical ways of selecting a right $h$ in this framework, as \cite{Hall13} deplored.

\ppn This paper fills this gap, as a sensible way of selecting such a value of $h$ is devised. A practical procedure, it does not claim to return an `undersmoothed' bandwidth or otherwise. It just aims to return a numerical value of $h$ which guarantees, for the data at hand, the intervals (\ref{eq:wald_kernel}), (\ref{eqn:wilson_interval_alt}) or (\ref{eqn:ACCIcond}) to have high degree of coverage accuracy. It is essentially a bootstrap procedure, which in many ways resembles \cite{Hall13}'s idea. A main difference, though, is that \cite{Hall13} looked for the (higher) nominal confidence level that they should target for their intervals, so that their empirical versions have a coverage probability close to $1-\alpha$. Here, the approach is more direct as the constructive parameter $h$ is the only focus. 

\ppn The procedure goes as follows:
\begin{enumerate}
 \item From the initial sample $\Xs$, estimate $p(x)$ by $\hat{p}_{h_0}(x)$ using an appropriate bandwidth $h_0$ returned by any data-driven procedure;
 \item Generate a large number $B$ of bootstrap resamples $\Xs^{*(b)}=\{(X_i,Y_i^{*(b)}); i=1,\ldots,n\}$, $b=1,\ldots, B$, according to $Y_i^{*(b)} \sim \text{Bernoulli}(\hat{p}_{h_0}(X_i))$;
 \item For $b \in \{1,\ldots,B\}$, compute on $\Xs^{*(b)}$ a collection of intervals\footnote{Here $CI(x;h)$ denotes a generic confidence interval for $p(x)$, which can be $CI_{\text{Wa}}(x;h)$, $CI_{\text{Wi}}(x;h)$ or $CI_{\text{AC}}(x;h)$.} $CI^{*(b)}(x;h)$ on a fine grid of candidate values of $h$;
 \item Estimate the coverage probability of the interval $CI(x;h)$ by the fraction $\widehat{P}(x;h)$ of intervals $CI^{*(b)}(x;h)$ which contain the `true' value $\hat{p}_{h_0}(x)$;
 \item Select for bandwidth one of the values of $h$ for which $\widehat{P}(x;h)$ is above $1-\alpha$. If $\widehat{P}(x;h)$ nowhere takes a value higher than $1-\alpha$, choose $h$ which maximises $\widehat{P}(x;h)$.
\end{enumerate}

\ppn The intervals (\ref{eq:wald_kernel}), (\ref{eqn:wilson_interval_alt}) and (\ref{eqn:ACCIcond}) heavily depend on the local geometry of the data around $x$, through the `local equivalent sample size' $nh\hat{f}_h(x)/R(K)$ (see comments below (\ref{eqn:fhat})). In order for the bootstrap intervals to mimic this appropriately, the resampling in 2.\ is done conditionally on the design values $\{X_i\}$ and those are kept fixed; only the $Y_i$'s are resampled. Also, theoretical considerations about bootstrap methods for nonparametric curve estimation suggest that resampling such as in 2.\ should be done from an estimate $\hat{p}_g$ with $g$, this time, an {\it oversmoothed} bandwidth -- with the same caveat about what this really means in practice. Again, the reason behind this is to do with the bias of the estimator, see e.g.\ \cite{Rodriguez93}. As opposed to other procedures, though, here the bootstrap resamples are only used for identifying another bandwidth, not for direct estimation of quantities of interest. Hence, oversmoothing is not theoretically required (see \cite{Hall13} for justification of this).

\ppn This procedure is tested through simulations in the next section, where it is seen to perform very well. Those simulations show that $\widehat{P}(x;h)$ is essentially a concave function of $h$, up to some minor fluctuations due to sampling (see Figure \ref{fig:hci} below), and that in almost all situations $\widehat{P}(x;h)$ indeed takes values higher than $1-\alpha$. Because of concavity, the values of $h$ for which it is the case forms a convex subset of $\R^+$. The value of $h$ chosen in 5.\ can then be the average of those values, for instance. This guarantees the selected value of $h$ to correspond to a value of $\widehat{P}(x;h)$ close to its maximum, hence leaning more to the side of conservatism than otherwise. For the rare cases in which $\widehat{P}(x;h)$ does not go above $1-\alpha$, its maximum value is very close to $1-\alpha$, anyway.

\subsection{Simulation study} \label{subsec:simstud}

In order to test, compare and validate the above construction of confidence intervals for $p(x)$ and the suggested bandwidth selection procedure, a twofold simulation study was run. 

\begin{figure}[h]
\centering
\includegraphics[width=0.4\textwidth]{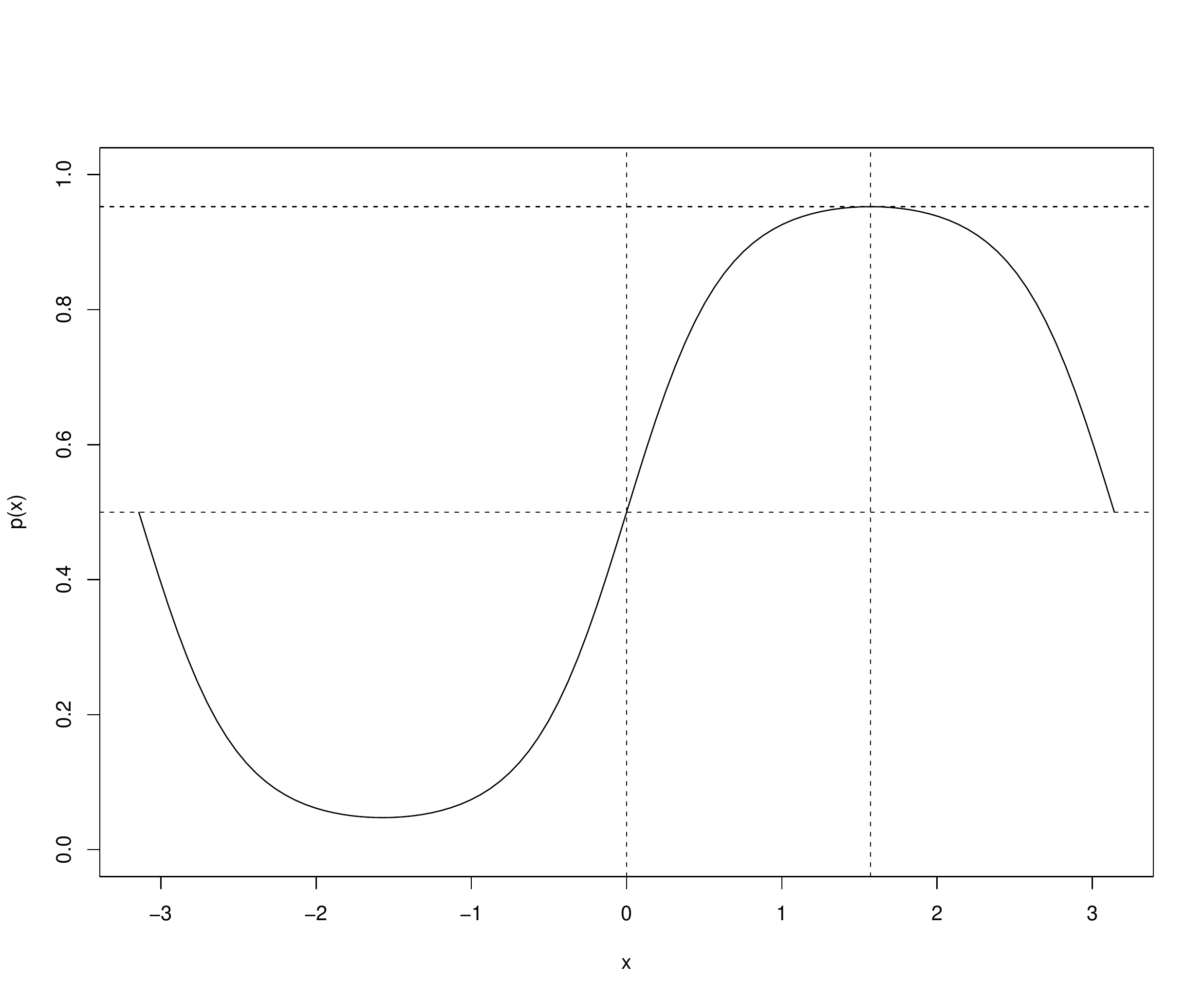}
\caption{Scenario 1: regression function $p(x) = \frac{e^{3 \sin(x)}}{1+e^{3 \sin(x)}}$. Dashed lines show the values of the function at $x=0$ and $x=\pi/2$.}
\label{fig:pscen1}
\end{figure}

\ppn {\bf Scenario 1.} The data was generated according to the following process:
\begin{align*}
 X & \sim \Us(-\pi,\pi), \\
Y | X & \sim  \text{Bernoulli}(p(X)), \\
p(x) &= \frac{e^{3 \sin(x)}}{1+e^{3 \sin(x)}}.
\end{align*}
This regression function, shown in Figure \ref{fig:pscen1}, was used in Example 5.119 in \cite{Wasserman2006}. The three confidence intervals (\ref{eq:wald_kernel}), (\ref{eqn:wilson_interval_alt}) or (\ref{eqn:ACCIcond}) for $p(x)$ at $x=0$ and $x= \pi/2$ were computed on $M=1,000$ independent samples generated as above. All confidence intervals were truncated to $[0,1]$ when necessary. The coverage probabilities of the three intervals, at the two locations, were approximated by the fraction of those $M=1,000$ intervals which include the true values $p(0) = 1/2$ and $p(\pi/2) \simeq 0.953$. In the non-conditional case, values of $p$ close to 0 or 1 are known to be problematic (see comments below (\ref{eqn:biasW})), so comparing how much impact the value of $p(x)$ has on the performance of the `conditional' intervals is of interest. To isolate that effect, a uniform design for $X$ was considered, to ensure that the areas `close to $x=0$' and `close to $x=\pi/2$' are equally populated by data.

\ppn Three sample sizes were considered: $n=50$ (`small' sample), $n=250$ (`medium' sample) and $n=1000$ (`large sample'). The targeted confidence level was 95\%, i.e.\ $\alpha = 0.05$. For each sample, the values of $h$ to use in (\ref{eq:wald_kernel}), (\ref{eqn:wilson_interval_alt}) or (\ref{eqn:ACCIcond}) were determined by the procedure described in Section \ref{subsec:USh}. The initial value of $h_0$ was taken here as the theoretically optimal value of the bandwidth for estimating $p$ - in this simulation study it is accessible as we know the `truth': it is $h_0 \simeq 0.745 n^{-1/5}$. This allows a fair comparison of the observed results as they are not impacted by other arbitrary decisions. The number of bootstrap replications was set to $B=1,000$ and the best value of $h$ was looked for on a grid of 200 equispaced values from 0.05 to 2. The final value of $h$ was taken as the centre (average) of the set of values producing an estimated coverage higher than 95\%, as suggested at the end of Section \ref{subsec:USh}. The (approximated) coverage probabilities of the intervals built according to this procedure are given in Table \ref{tab:simCP}.

\begin{table}[H] \centering
 \begin{tabular}{||l l || c | c || }
\hline \hline
 & & $x=0$ & $x=\pi/2$ \\ \hline 

$n=50$ & Wald &  0.862 & 0.939 \\
 & Wilson & 0.971 & 0.907 \\
 & Agresti-Coull & 0.974 & 0.920 \\
\hline

$n=250$ & Wald &  0.908 & 0.796  \\
 & Wilson &  0.932 & 0.940 \\
 & Agresti-Coull & 0.944  & 0.939 \\
\hline 

$n=1000$ & Wald & 0.933 & 0.860 \\
 & Wilson & 0.952 &  0.958 \\
 & Agresti-Coull & 0.951  & 0.961 \\
\hline \hline
\end{tabular}
\caption{Scenario 1, (approximated) coverage probabilities for the three types of confidence intervals (nominal confidence level: 95\%) at $x=0$ and $x=\pi/2$, for sample sizes $n=50$, $n=250$ and $n=1000$.} \label{tab:simCP}
\end{table}

\ppn It emerges from Table \ref{tab:simCP} that the Wilson and Agresti-Coull intervals reach an empirical coverage consistently very close to the targeted 95\%, and this for all sample sizes and at both locations (i.e., both when $p(x) \simeq 1/2$ and $p(x) \simeq 1$). A notable exception, though, is when $x = \pi/2$ and $n=50$ (`small' sample). As explained in Section \ref{subsec:WaldCondCi}, the number of observations effectively playing a role in estimating $p$ at $x$ is $nh \hat{f}_h(x)/R(K)$. Here, with $n=50$, $f \equiv 1/(2\pi)$ and $R(K) = 1/(2\sqrt{\pi}$) (for $K=\phi$ the standard Gaussian kernel), the local equivalent sample size is roughly 1 observation for the values of $h$ around $h_0$. That means that the effect of the shrinkage described below (\ref{eqn:wilson_interval_alt}) is severe here, and the centre of the interval is seriously held back toward 1/2. At $x=0$, this is a good thing as $p(0)=1/2$, and the observed coverage is actually higher than 95\%; at $x=\pi/2$, with $p(\pi/2)$ close to 1, this is detrimental to the level of the intervals (which keep an empirical coverage higher than 90\%, though). As expected given the behaviour of its non-conditional counterpart, the Wald interval struggles to maintain a reasonable level of coverage, even at large sample sizes or in favourable cases ($x=0$). Only when $x=0$ and $n=1000$ does the Wald interval produces reasonable results (but still slightly less accurate in terms of coverage probability than Wilson and Agresti-Coull). 

\begin{figure}[h]
\centering
\includegraphics[width=0.6\textwidth]{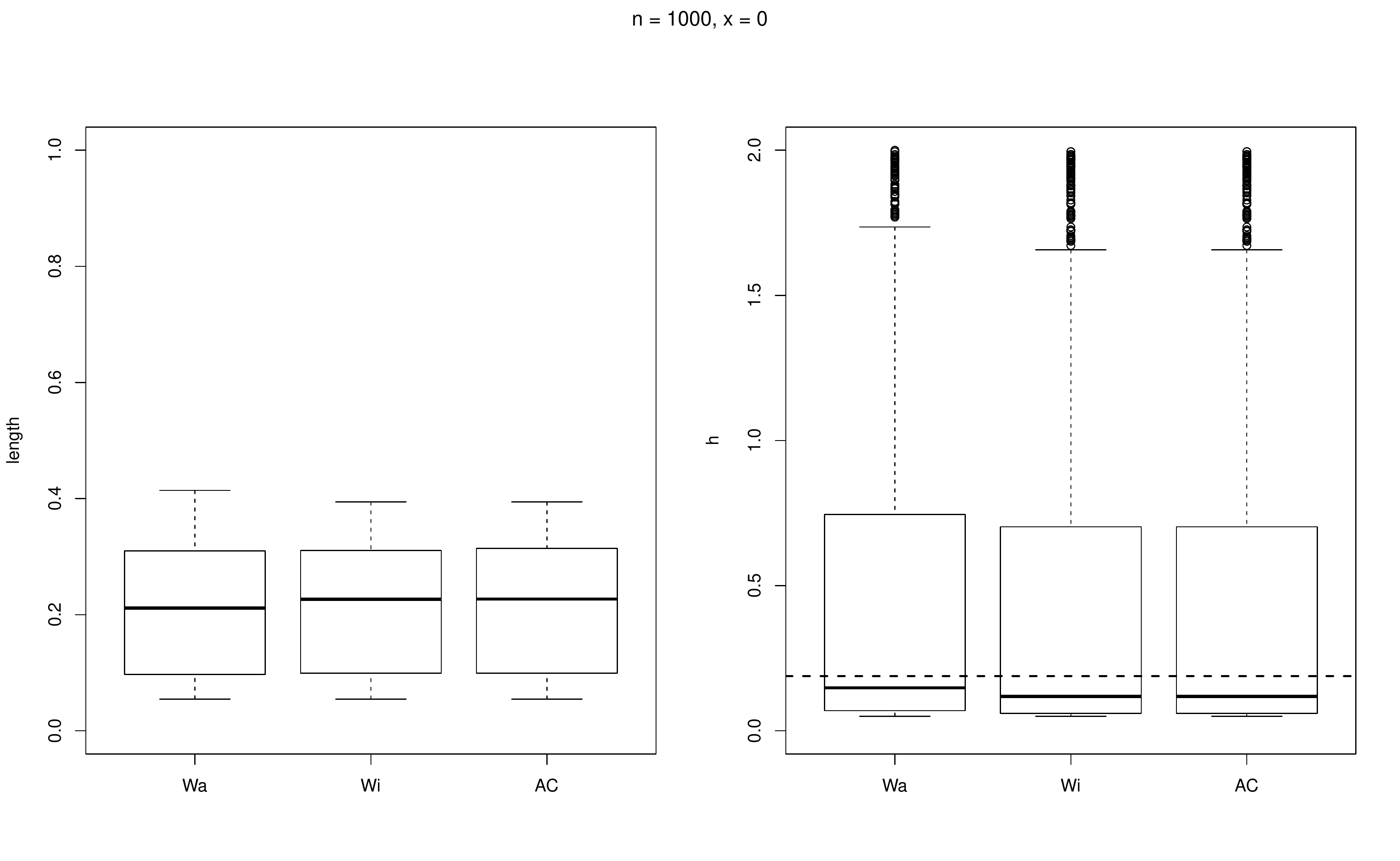}
\caption{Scenario 1: length of the intervals (left) and selected values of $h$ (right) for the three types of confidence intervals computed at $x=0$ on $M=1,000$ independent samples of size $n=1000$. The dashed line (right panel) shows $h_0 = 0.118$.}
\label{fig:lengthh0}
\end{figure}

\begin{figure}[h]
\centering
\includegraphics[width=0.6\textwidth]{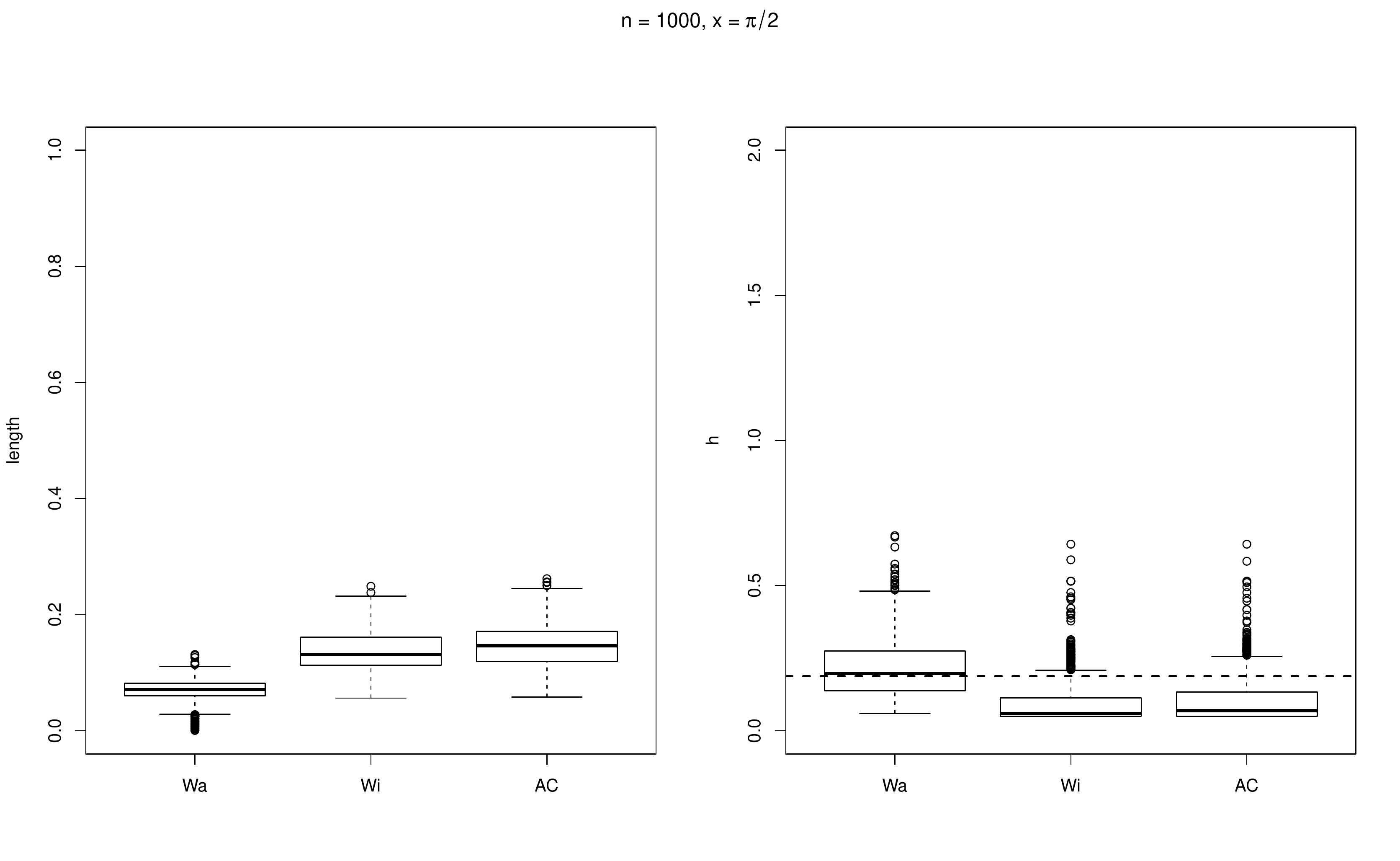}
\caption{Scenario 1: length of the intervals (left) and selected values of $h$ (right) for the three types of confidence intervals computed at $x=\pi/2$ on $M=1,000$ independent samples of size $n=1000$. The dashed line (right panel) shows $h_0 = 0.118$.}
\label{fig:lengthhpi}
\end{figure}

\ppn Another perspective on this is provided by Figures \ref{fig:lengthh0} and \ref{fig:lengthhpi}, which show boxplots of the lengths of the $M=1,000$ confidence intervals computed from each construction for $n=1000$, as well as the selected values of $h$, for $x=0$ and $x=\pi/2$. At $x=0$, the three intervals are always very similar, and so are the values of $h$ selected by the procedure described in Section \ref{subsec:USh}. The empirical coverage are, therefore, similar as well as shown by Table \ref{tab:simCP}. At $x=\pi/2$, however, the procedure selects values of $h$ much smaller for the Wilson and Agresti-Coull intervals, than for the Wald interval (Figure \ref{fig:lengthhpi}, right panel). This means that the constructed Wilson and Agresti-Coull intervals are indeed longer than the Wald interval (Figure \ref{fig:lengthhpi}, left panel), but that is the price to pay to keep a coverage probability of 95\%. Recalling that $h$ is selected via a bootstrap procedure, the value of $h$ guaranteeing a high coverage for the bootstrap replications of the Wald interval, is not guaranteed to maintain such high coverage `in the real world'. For the Wilson and Agresti-Coull intervals, on the other hand, that is the case.

\ppn Figures \ref{fig:lengthh0} and \ref{fig:lengthhpi} also show the optimal value $h_0 = 0.188$ (for $n=1000$; dashed line in the right panel). According to the boxplots, the value $h$ supposed to be good for constructing the confidence intervals (\ref{eq:wald_kernel}), (\ref{eqn:wilson_interval_alt}) or (\ref{eqn:ACCIcond}) is, for many samples, smaller than $h_0$, in agreement with what `undersmoothing' suggests. It is, however, {\it not always} the case. Oftentimes (especially at $x=0$), taking $h$ (much) greater than $h_0$ seems to be the right thing to do. The fact that a `small' $h$ is not always ideal is easily understood through the case $x=0$. In this scenario, due to symmetry, $p(0)$ is actually equal to $\E(Y) = p$, the non-conditional probability $\P(Y=1)$. As a result, any confidence interval for $p$ such as those described in Section \ref{sec:chap2} can be used for $p(0)$ as well. This is advantageous, as the sample average $\bar{Y}=\hat{p}$ is naturally a better estimator (smaller variance, no bias) of the global $p$ than any local ($h$ `small'), conditional attempt. Those `non-conditional' intervals are actually recovered from (\ref{eq:wald_kernel}), (\ref{eqn:wilson_interval_alt}) or (\ref{eqn:ACCIcond}) as $h\to \infty$. Indeed, it is known that taking a large bandwidth in nonparametric regression essentially makes local estimators into global ones \citep{Eguchi03}. Intuitively, if $h$ is very large ($h \simeq \infty$) in (\ref{eqn:NWest}) then all observations are equally weighted and $\hat{p}_\infty(x)$ just reduces to the sample average $\bar{Y}=\frac{\YY}{n}=\hat{p}$. Therefore, it is beneficial to take $h$ `large' here. Of course, this is a very particular situation, but it exemplifies that the best $h$ really depends on the design and is not necessarily `small'. Hence heuristic rules such as taking $h = C_n h_0$, with $C_n$ a small constant (possibly depending on $n$), are thus to be precluded and $h$ should be selected by a careful data-driven selection procedure. This shows the value of the procedure developed in Section \ref{subsec:USh}.

\begin{figure}[h]
\centering
\includegraphics[width=0.6\textwidth]{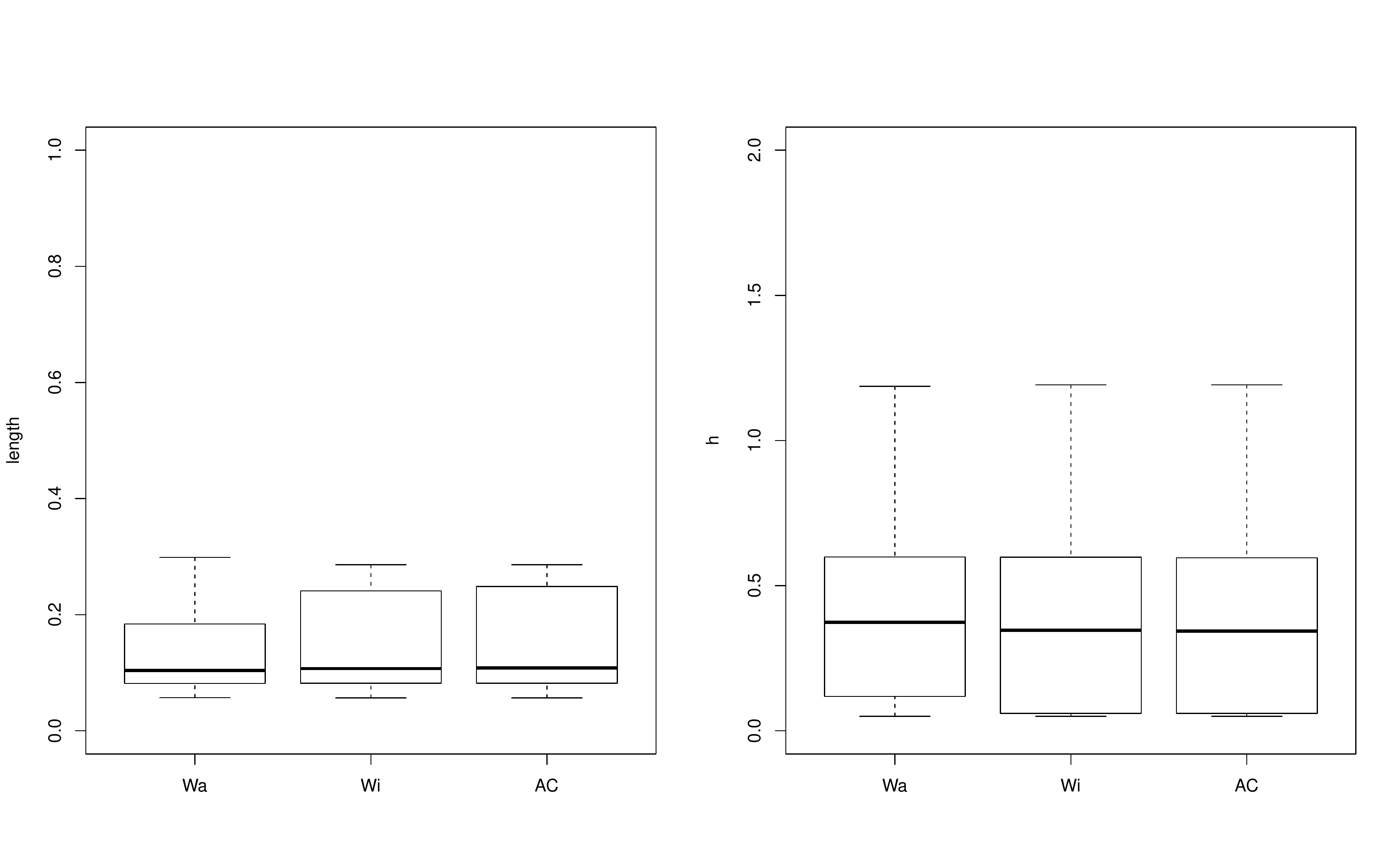}
\caption{Scenario 2: length of the intervals (left) and selected values of $h$ (right) for the three types of confidence intervals computed at $x=0$ on $M=1,000$ independent samples of size $n=1350$.}
\label{fig:lengthh2}
\end{figure}

\ppn {\bf Scenario 2.} The purpose of this second scenario is to empirically validate the real data analysis shown in the next section. Essentially the same study as in Scenario 1 was repeated, but this time data sets of size $n=1350$ were generated as
\begin{align*}
 X & \sim 0.45 \times \Ns(-1,1/4) + 0.55 \times \Ns(0.8,1/4), \\
Y | X & \sim  \text{Bernoulli}(p(X)), \\
p(x) &= \frac{1}{1+e^{-0.088-0.770 x}}.
\end{align*}
The above mixture of Normals is a good parametric approximation of the distribution of the predictor $X$ in the application below (Figure \ref{fig:fhat}), the above function $p$ is the best logistic fit for the analysed data (Figure \ref{fig:phatlogit}), and the sample size $n=1350$ is akin to that sample size as well. Hence the results of this simulation gives an appreciation of the validity of the real case analysis described in the next Section. Again, $M=1,000$ independent samples were generated. For each of them, the values of $h$ to use in (\ref{eq:wald_kernel}), (\ref{eqn:wilson_interval_alt}) or (\ref{eqn:ACCIcond}) were determined by the procedure described in Section \ref{subsec:USh}. The coverage probabilities of the three types of confidence intervals for $p(0)$ was approximated by the fractions of the $M=1,000$ such intervals which include the true $p(0) = 0.522$. Those were $0.934$ for the Wald interval, $0.953$ for the Wilson interval, and $0.955$ for the Agresti-Coull interval. Figure \ref{fig:lengthh2} shows the lengths and selected values of $h$ for the $M=1,000$ independent samples generated in this scenario. The conclusion are very similar to what was said for the case $x=0$, $n=1000$ in Scenario 1. In particular, both the Wilson and Agresti-Coull intervals show coverage probabilities very close to their nominal level 95\%, while, as $p(0) \simeq 1/2$ and $n$ is `large', the Wald interval is not doing too bad either. This indicates that the conclusions drawn in Section \ref{subsec:anal} can be given some credibility.

\section{Analysis of UEFA competitions} \label{sec:chap4}

\subsection{UEFA coefficients} \label{subsec:background}

The main aim of this analysis is to investigate the existence, or otherwise, of the Second Leg Home Advantage (SLHA) in UEFA competitions (Champions League and Europa League). The SLHA is described as the advantage in a balanced two-legged knockout tie whereby teams are, on average, more likely to qualify if they play at home in the second leg. Hereafter, the team playing at home on the second leg will be called `Second Leg Home Team' [SLHT], and conversely for the `First Leg Home Team' [FLHT]. Table 1 in \citet{Page2007a} exemplifies that many players and coaches believe that the SLHA exists. Maybe more surprising is that the UEFA body itself seems to believe in it as well: the design of the UEFA Champions and Europa Leagues is such that teams which finished first in their group in the first stage, are {\it rewarded} by being guaranteed to play the second leg at home in the first subsequent knockout round. 

\ppn If only for that, any serious empirical analysis of the SLHA effect must adjust for team strengths, as briefly explained in Section \ref{sec:intro}. This said, the `strength' of a team is a rather vague concept, which in addition is likely to vary over the course of a season according to many imponderable factors (e.g., the injury of a key player may seriously affect a team's abilities). A simple proxy for such `strength' seems to be the UEFA club coefficient. Its calculation is actually based on two elements: first, an individual club index, obtained from the points awarded to the team for their performance during the course of the Champions or Europa league each year (details to be found in \cite{Kassies}); second, a country index, which is the average of the indices of all the clubs of the same country which took part to a European competition that year. The country index for a particular season is defined as the sum of that country's points from the previous five seasons. Since 2009, the club UEFA coefficients have been calculated as the sum of the clubs points from the previous five seasons plus the addition of 20\% of the relevant country coefficient for the latest season.\footnote{Between 2004-05 and 2008-09, the country weight was 33\%, whereas it was 50\% before 2003-04. \label{foot:coutnry}} The club coefficients take thus into account the recent performance of the club in European competitions, but also the strength of the league from which the team comes. Although not perfect, they form the basis for ranking and seeding teams in the UEFA competitions, and should be reasonably representative of the relative overall strengths of the teams over one season.

\ppn Support for this assertion is indirectly provided in \cite{Eugster2011}. They hypothesised that, as the club coefficients are only updated at the end of each season, group stage performance would provide a more up-to-date measure of team strength for the following knockout rounds. As such, they included both club coefficients and group stage performance as predictors in their logistic regression model. It turned out that `group stage performance' was not statistically significant either in a model by itself or in a model which also used `club coefficient' (which was significant). It seems, therefore, fair to consider the UEFA coefficient of clubs as the main indicator of team strength for a given season, and the modelling below will make use of that index only.

\subsection{Preliminaries} \label{subsec:dataprep}

\ppn {\bf Data preparation.} The analysed data, all drawn from \cite{Kassies}, include all match results from UEFA Champions and Europa leagues from 2009/2010 through to 2014/15 and the UEFA coefficients of all clubs which took part to some European competition for those seasons. Initially, data from seasons 1999/00 to 2014/15 were planned to be analysed, however UEFA has changed the method for calculating the club coefficients three times during that period. These changes mean that the time periods (1999/00 to 03/04, 04/05 to 08/09, and 09/10 onwards) are no longer directly comparable given that the UEFA coefficients would be fundamentally different variables. This necessitated analysis of only one time period, and the most recent (2009/10 to 2014/15) was naturally chosen. It would have been possible to recalculate the UEFA coefficients of each club for the years prior to 2009 using the current method, but it is probably meaningful here to focus on the last years only, given the possible fading of the SLHA over years that have sometimes been suggested by previous studies (see Section \ref{sec:intro}). This analysis will consequently evaluate the existence of the SLHA `now', and not from a historical perspective.

\ppn Altogether, 4160 matches were played in the Champions and Europa leagues from 2009/2010 to 2014/2015. Of course, only the knockout two-legged ties were of interest in this study. Hence, the first, obvious action was to remove group stage games, played in a round robin style and not qualifying as a two-stage knockout, and match-ups where only one game was played such as Finals or match-ups where one game was cancelled (it happened that some games were cancelled due to security concerns, or others\footnote{{\tt http://www.uefa.org/news/newsid=1666823.html}}). It remained $n=1353$ two-legged ties (i.e., pairs of matches), both in qualifying rounds (before the group stage) and in final knockout stage (after the group stage).

\ppn In a two-legged knockout tie, the aggregate score over the two matches is tallied and the team which scored the most goals on aggregate is declared the winner and qualifies to the next round. If the teams have scored an equal number of goals, the so-called `{\it away goals rule}' applies: the team which scored most `away from home' would qualify. If this criterion does not break the tie, then two extra periods of 15 minutes are played at the end of the normal time of the second leg. If the teams are still tied then, the result is decided by penalty shoot-out. For games which went to extra-time (there were $n_\text{ET} = 84$ of them), it is reasonable to hypothesise that some SLHA may be induced by the fact that the SLHT plays 30 minutes longer at home than the FLHT, hence benefiting more from its home advantage \citep{Nevill99}. This could justify to exclude those games from the study, given that they {\it should} artificially give rise to some sort of SLHA. It was, however, decided to keep them in a first time, arguing that it might precisely be the key element to take into account when assessing the SLHA. The possibility of playing extra time and/or taking penalties on their home ground might explain why most players favour playing at home on the second leg. In a second time, those $n_\text{ET}=84$ games will be excluded from the analysis, in order to appreciate the real effect of extra-time/shoot-out on the SLHA (see Section \ref{subsec:disc}).

\ppn {\bf Predictor.} Call $Y$ the binary outcome of a two-legged tie, and for the $i$th tie define $Y_i = 1$ if the SLHT qualifies and $Y_i = 0$ otherwise (hence, if the FLHT qualifies). The study is based on the regression of $Y$ on an explanatory variable $X$ quantifying the inequality in strength between the two teams involved. Call $C_1$ and $C_2$ the UEFA coefficients of the FLHT and SLHT, respectively, at the time of meeting, and define\footnote{There are actually two occurrences where teams had a coefficient of zero, which causes a problem for defining their logarithm: two teams from Gibraltar made their first appearances in 2014/15, and Gibraltar itself was a newly accepted member of the UEFA so had a country coefficient of zero as well. Those teams were artificially given a coefficient of $C=0.001$, well below the next smallest coefficient of 0.050.} 
\begin{equation} X = \log(C_2/C_1)=\log(C_2) - \log(C_1). \label{eqn:predX} \end{equation}
A positive value of $X$ indicates that the SLHT is stronger than the FLHT, and conversely for a negative $X$. The value $X=0$ indicates an exactly balanced tie. The value $X_i$ is the observed value of $X$ for the $i$th tie.

\ppn Note that \cite{Page2007a} and \cite{Eugster2011} used the difference $C_2-C_1$ as control variable (\cite{Eugster2011} actually used a normalised version of it). The reason why a log transformation is introduced in (\ref{eqn:predX}) is that $C_1$ and $C_2$ are positive variables. It turns out that two positive numbers are more naturally compared through their ratio than through their difference.\footnote{There are mathematical reasons for this. Algebraically, $(\R^+,\times)$ is a group, $(\R^+,+)$ is not. The Haar measure (i.e., the `natural' mathematical measure) on $(\R^+,\times)$ is $\nu(dx) = \frac{dx}{x}$, translating to the measure of an interval $[a,b] \subset \R^+$ being $\log(b)-\log(a)$.} As an illustration, in the first qualifying round of the Champions League 2014-2015, the Sammarinese team of La Fiorita ($C_1=0.699$) faced the Estonian team of Levadia Tallinn ($C_2=4.575$); the same year, in the semi-finals, FC Barcelona ($C_1=157.542$) clashed with Bayern M\"unchen ($C_2 = 154.328$). The (absolute) difference in coefficients is roughly the same for both ties ($3.876$ and $3.214$, respectively), however anybody with a slight appreciation for European football would know that the second case, bringing together two giants of the discipline, was to be much tighter than the first one, opposing a new-coming team from one of the weakest leagues in Europe (San Marino) to a more experienced team from a mid-level league. The ratio of the coefficients $C_2/C_1$, respectively $6.545$ and $1.021$ for the above two ties, is much more representative of the relative forces involved. 

\begin{figure}[h]
\centering
\includegraphics[width=0.5\textwidth]{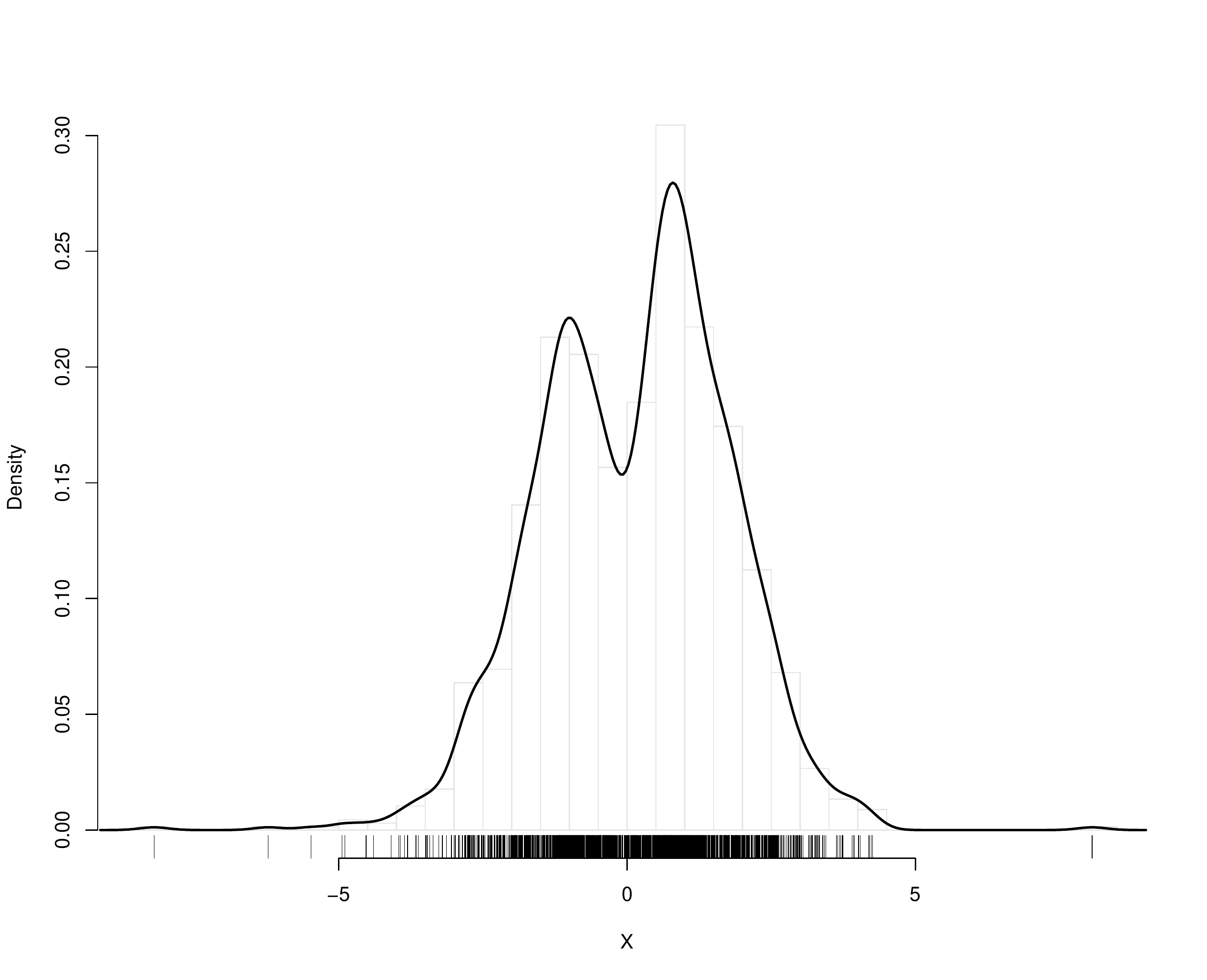}
\caption{Kernel estimate (\ref{eqn:fhat}) of $f$, with $h=0.252$ and $K=\phi$ the standard Gaussian density. The ticks at the bottom of the graph show the observed values $X_i$'s.}
\label{fig:fhat}
\end{figure}

\ppn Figure \ref{fig:fhat} shows the kernel estimate $\hat{f}$ (\ref{eqn:fhat}) of the density $f$ of the predictor $X$, overlaid to an histogram. The bandwidth $h=0.252$ was selected by direct plug-in \citep[Section 3.3.3]{Hardle2004} and the kernel was the standard Gaussian density. The estimate clearly suggests that $f$ is bimodal. This can be understood the following way. In the qualifying rounds, it is very rare to see two teams of very similar strength facing each other: there is often a team `much stronger' (at that level) than the other. See the above example La Fiorita versus Levadia Tallinn: although the two teams can be considered `weak' and both their UEFA coefficients are `low', the ratio of those coefficients unequivocally tells which team is likely to qualify. A value of $X$ close to 0 is actually only observed if the UEFA coefficients of the two matched-up teams are {\it really of the same order}, and that typically happens in the final rounds, when the strongest teams meet (for example, FC Barcelona versus Bayern M\"unchen, see above). There are, obviously, much more qualifying games than semi-finals, hence there are comparatively less games characterised by a value of $X$ close to 0, than otherwise. In any case, $\hat{f}$ shows a peak on the positive side noticeably higher than that on the negative side. In fact, the observed proportion of positive values $X_i$'s is $752/1353 = 0.556$ (Wilson confidence interval: $[0.529;0.582]$). This means that the stronger team is indeed more often the SLHT than the contrary, which confirms the existence of the confounding factor described in Sections \ref{sec:intro} and \ref{subsec:background}, and the necessity of taking it properly into account.

\ppn {\bf Model.} In this application, it is is sensible to treat the observed values $X_i = \log(C_{2,i})-\log(C_{1,i})$ of the predictor as known constants set by the design (`{\it fixed design}'), and it will be assumed that
\[Y_i |X_i \sim \text{Bernoulli}(p(X_i)) \]
for $i = 1,\ldots, n =1356$, independently of one another. The function $p$ is left totally unspecified except that it is twice continuously differentiable.

\subsection{Analysis} \label{subsec:anal}

\begin{figure}[h]
\centering
\includegraphics[width=\textwidth]{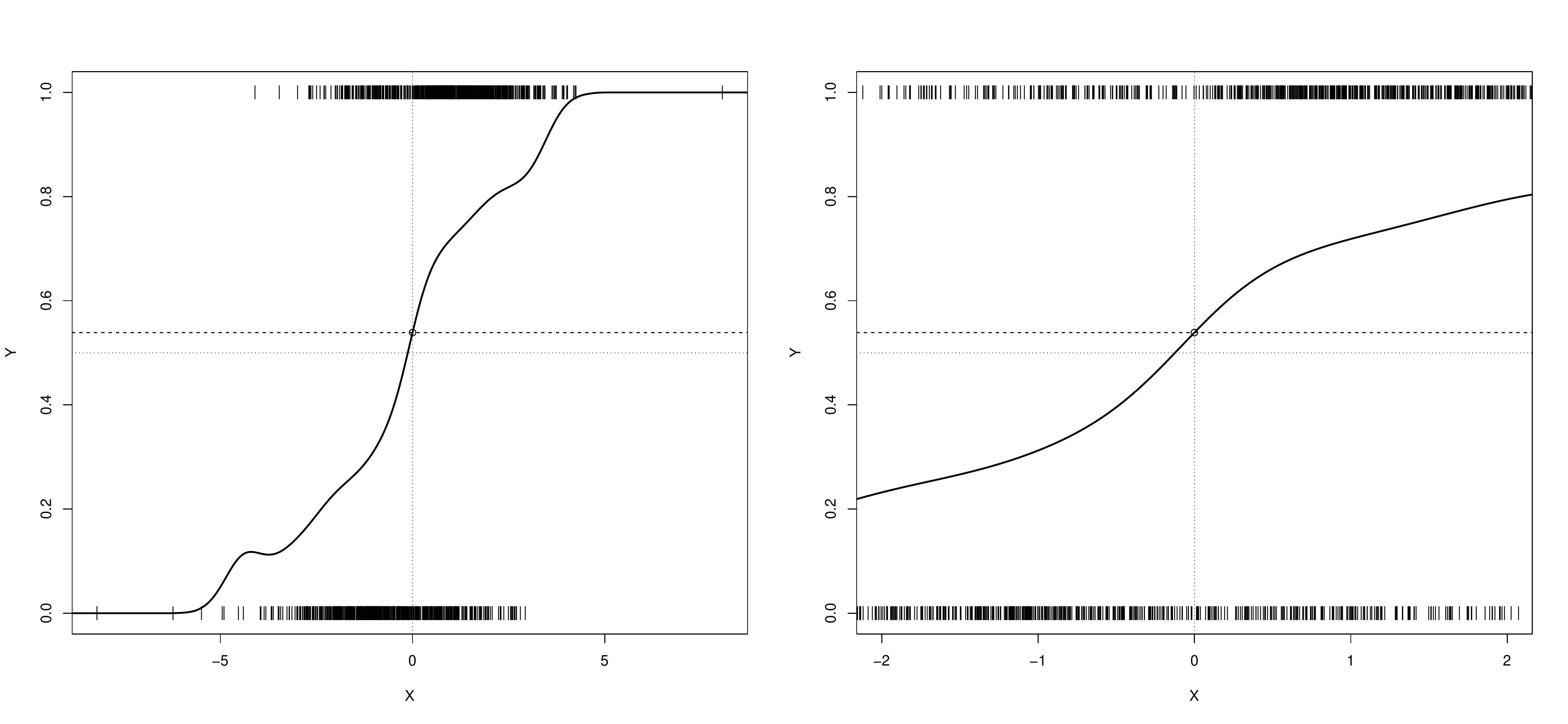}
\caption{Nadaraya-Watson estimator $\hat{p}_{h_0}$ (\ref{eqn:NWest}) of $p$, with $h_0=0.525$ and $K=\phi$ the standard Gaussian density. Global view (left) and close-up around $x=0$ (right). Ticks show the observed data.}
\label{fig:phat}
\end{figure}

The Nadaraya-Watson estimator $\hat{p}_{h_0}$ (\ref{eqn:NWest}) was computed on the data set $\Xs = \{(X_i,Y_i); i=1,\ldots,n\}$. The kernel $K$ was the standard Gaussian density, while the optimal bandwidth was approximated by the method based on AIC described in \cite{Hurvich98},\footnote{This bandwidth selector is implemented in the R package {\tt np}.} which returned $h_0 = 0.525$. The resulting estimate is shown in Figure \ref{fig:phat} (left). Common sense suggests that $p$ should be a monotonic function of $x$, hence the little `bump' in $\hat{p}_{h_0}$ between $x=-5$ and $x=-4$ is most certainly due to random fluctuation only. This happens in an area where data are rather sparse, so it is not surprising that the nonparametric, essentially local, estimator is not the most accurate there. If the focus of the analysis was that part of the range of values of $X$, then more involved estimation methods could be used (e.g., adaptive estimation based on variable bandwidths, robust version of the NW estimator such as LOESS, or isotonic regression). However, this study is mainly interested by what happens around $x=0$, where data are abundant. The NW estimate $\hat{p}_{h_0}$ looks well-behaved and smooth over $[-2,2]$, say (see close-up in Figure \ref{fig:phat}, right), so it is probably enough here. In particular, the estimator gives $\hat{p}_{h_0}(0) = 0.539 > 1/2$, indicating a potential SLHA. The statistical significance of this effect is examined below by computing the confidence intervals (\ref{eq:wald_kernel}), (\ref{eqn:wilson_interval_alt}) and (\ref{eqn:ACCIcond}) for $p(0)$, using a bandwidth $h$ obtained from the procedure described in Section \ref{subsec:USh}. Details are given below.

\begin{figure}[h]
\centering
\includegraphics[width=1\textwidth]{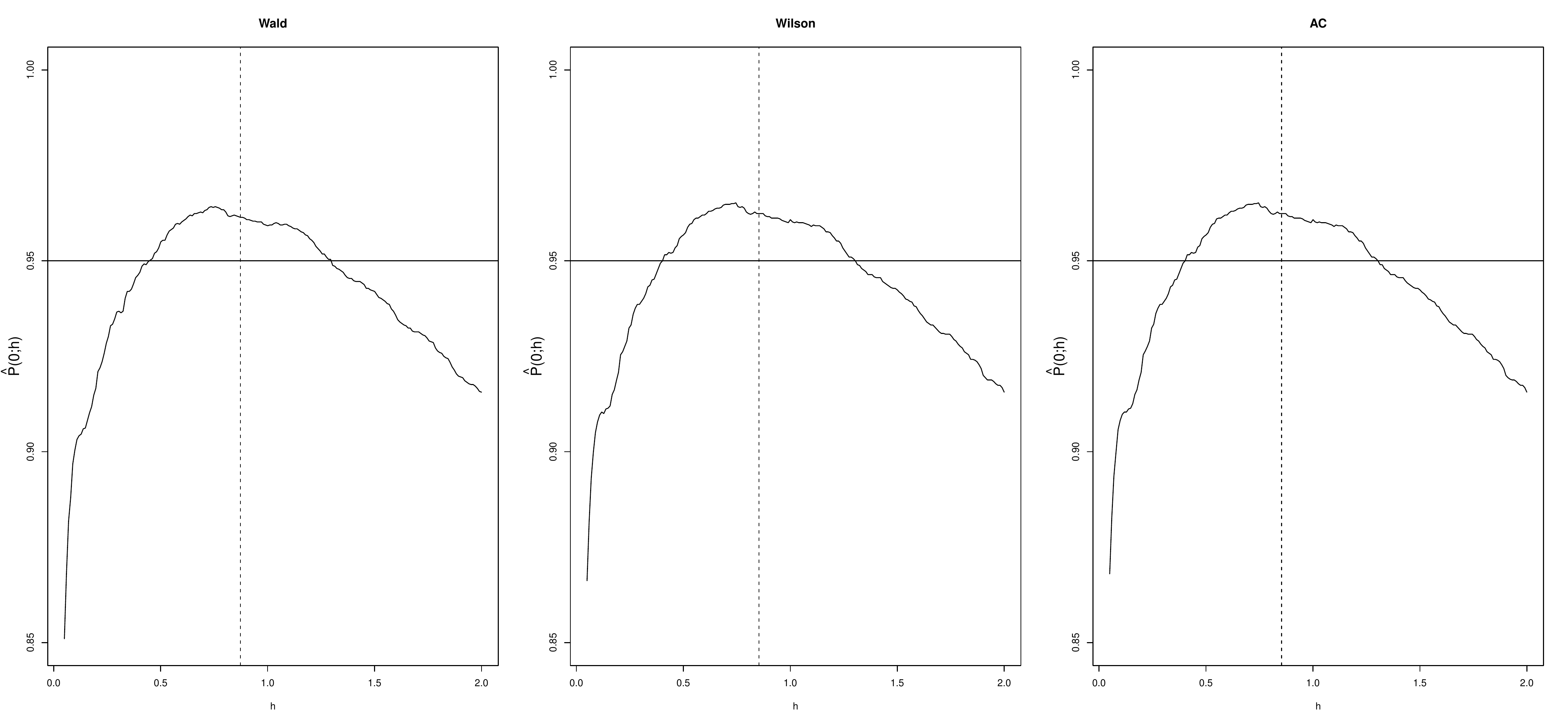}
\caption{Appearance of the functions $\widehat{P}(0;h)$ for the 3 types of confidence intervals. The dashed lines show the returned values of $h$, obtained as the average of the set of values for which $\widehat{P}(0;h)$ exceeds 95\%.}
\label{fig:hci}
\end{figure}

\ppn Step 1.\ is what Figure \ref{fig:phat} shows. In Step 2., $B=5,000$ bootstrap resamples were generated as $Y_i^{*(b)} \sim \text{Bernoulli}(\hat{p}_{h_0}(X_i))$; $i=1,\ldots,n$, $b = 1,\ldots,B$.  For Step 3., an equispaced grid of 200 candidate values for $h$, from $h=0.05$ to $h=2$, was built. On each bootstrap resample, the 3 confidence intervals (\ref{eq:wald_kernel}), (\ref{eqn:wilson_interval_alt}) and (\ref{eqn:ACCIcond}) at $x=0$ for each candidate value of $h$ were computed. The appearance of the functions $\widehat{P}(0;h)$ (Step 4.) for the three types of intervals is shown in Figure \ref{fig:hci}. The returned values of the bandwidth to use (Step 5.) were obtained as the average of all the values $h$ which give an estimated coverage higher than 95\%. Those were: $h=0.873$ for the Wald interval, and $h=0.854$ for both the Wilson and the Agresti-Coull interval. With these values of $h$ in (\ref{eq:wald_kernel}), (\ref{eqn:wilson_interval_alt}) and (\ref{eqn:ACCIcond}), we obtain as 95\% confidence intervals for $p(0)$:
\begin{equation} CI_{\text{Wa}}(x=0;h=0.873) = CI_{\text{Wi}}(x=0;h=0.854) = CI_{\text{AC}}(x=0;h=0.854)  = [0.504; 0.574]. \label{eqn:CIUEFA} \end{equation}
Interestingly, with this sample size, the three intervals are mostly indistinguishable, as they differ only from the fourth decimal digit. Of course $p=1/2$ does not belong to this interval, evidencing the statistical significance of the SLHA (more on this in Section \ref{subsec:disc}).

\ppn From a methodological point of view, it is interesting to note that the empirically `optimal' bandwidth $h$ for building the confidence intervals (here $h \simeq 0.86$) is actually {\it greater} than the bandwidth $h_0 = 0.525$, considered `best' for estimating $p$. As already noted in Section \ref{subsec:simstud}, this is in contrast to what a naive interpretation of `undersmoothing' would suggest (to take $h$ smaller than $h_0$). Admittedly, $h_0$ still belongs to the acceptable area ($\{h >0: \widehat{P}(0;h) \geq  0.95\}$), see Figure \ref{fig:hci}. However, if one took $h = h_0 n^{-1/3}/n^{-1/5}$ as it has sometimes been suggested in the literature following (\ref{eqn:covprob}), it is here $h=0.2$ and the so-produced intervals would have a coverage probability much lower than the targeted 95\%. Note that taking $h$ too small has also an adverse effect of the length of the intervals, given that the standard error of the estimator $\hat{p}_h(x)$ is essentially inversely proportional to $h$, by (\ref{eqn:asnormNW})-(\ref{eqn:NW_asymp_normal_h_undersmooth}). This confirms that {\it ad-hoc} procedures aiming at producing a supposedly `undersmoothed' bandwidth need not work well in finite samples, and should not be used.

\begin{figure}[h]
\centering
\includegraphics[width=0.4\textwidth]{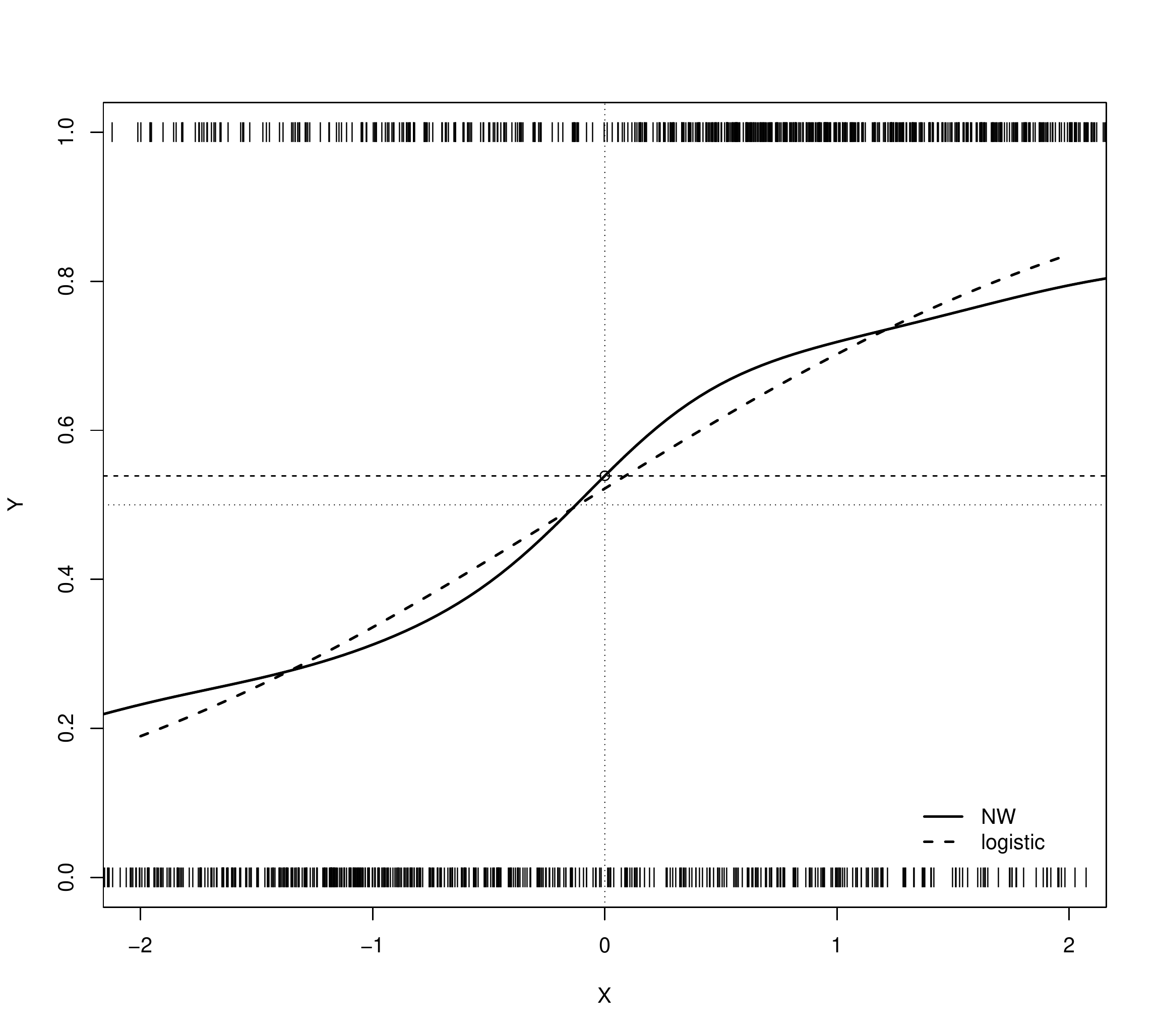}
\caption{Nadaraya-Watson estimator of $p$ (solid line, same as in Figure \ref{fig:phat}) and fitted logistic curve (dashed line).}
\label{fig:phatlogit}
\end{figure}

\ppn For completeness, some elements of analysis based on a parametric logistic model are briefly given below. Importantly, it is noted that a classical goodness-of-fit test for GLM based on the deviance and Pearson's residuals rejects the logistic model for the data ($p$-value $\sim 0.001$). The le Cessie-van Houwelingen test \citep{Cessie91}, precisely based on the Nadaraya-Watson estimator (\ref{eqn:NWest}), shows marginal evidence against it as well ($p$-value $=0.06$). Therefore, what follows is not fully supported by the data, and is shown for illustration only. The fitted logistic model is $\logit(p(x)) = \alpha + \beta x$, and the coefficients are estimated at $\hat{\alpha} = 0.088$ and $\hat{\beta} = 0.770$. This logistic fit is shown in Figure \ref{fig:phatlogit}, overlaid to the Nadaraya-Watson estimator, to appreciate their discrepancy around 0. A 95\% confidence interval for $\alpha$ is $[-0.035;0.210] \ni 0$, indicating the non-significance of the intercept. Given that $p(0) = e^\alpha/(1+e^\alpha)$, this translates into a 95\% confidence interval $[0.491,0.552] \ni 1/2$ for $p(0)$. Hence an analysis based on logistic modelling would fail to highlight a potential SLHA. Figure \ref{fig:phatlogit} reveals that the constrained logistic specification for $p$ forces the estimate to `take a shortcut' compared to what the data really say (i.e., the nonparametric estimate), and that smoothes over the interesting features around 0. Remarkably, the length of the interval for $p(0)$ based on this parametric model is $0.061$, whereas the nonparametric confidence interval (\ref{eqn:CIUEFA}) is only slightly longer (margin of $0.070$). The price to pay for the flexibility and robustness granted by the nonparametric approach does not seem to be in terms of precision (i.e., length of the confidence intervals) in this study.

\begin{figure}[h]
\centering
\includegraphics[width=0.4\textwidth]{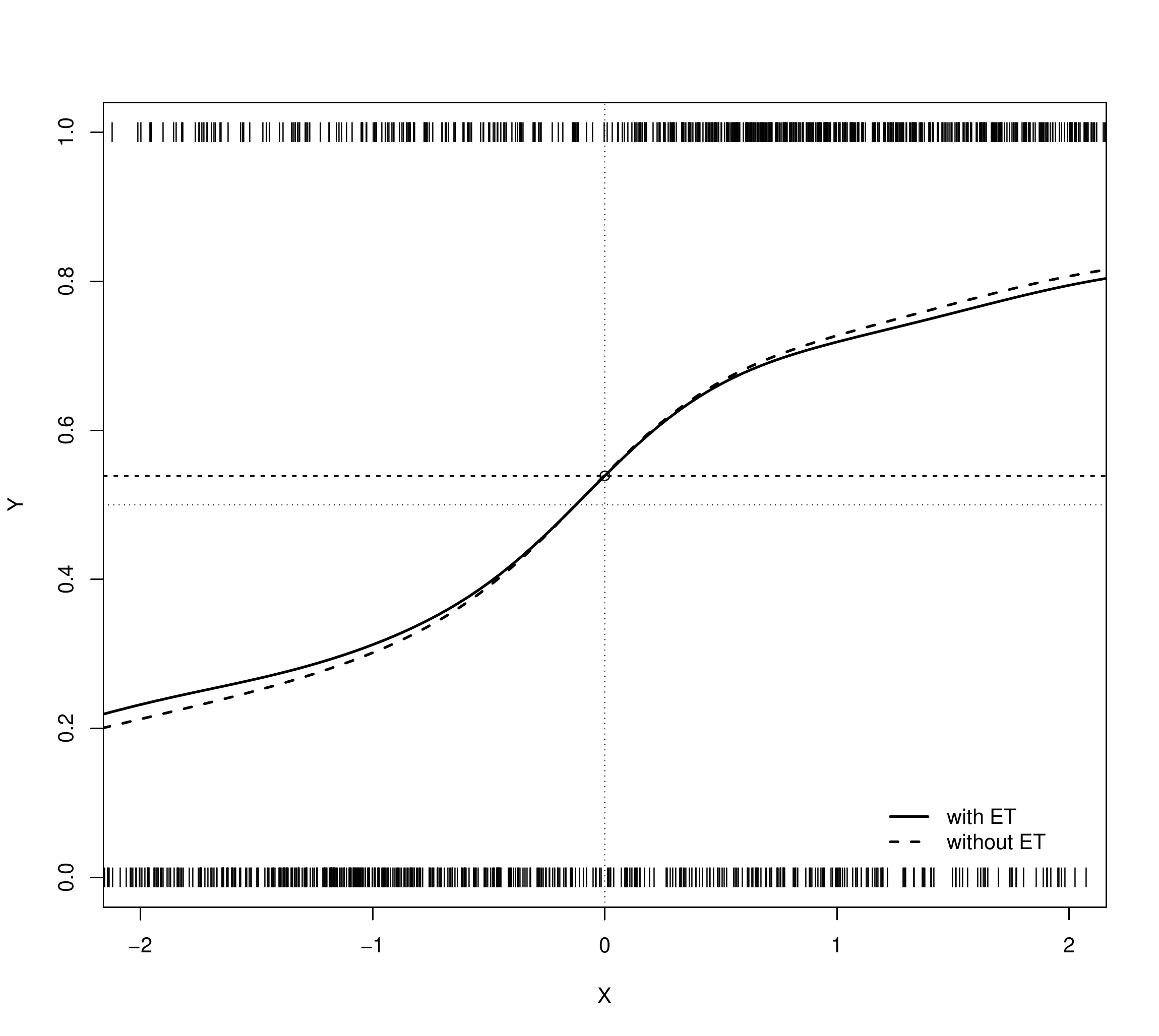}
\caption{Nadaraya-Watson estimator of $p$ (solid line, same as in Figure \ref{fig:phat}) and fitted logistic curve (dashed line).}
\label{fig:phatET}
\end{figure}

Finally, the nonparametric analysis was repeated, but this time excluding the ties which went to extra-time, in order to appreciate the effect of those on the existence or otherwise of the SLHA. There were $n_\text{ET} = 84$ such occurrences of extra-time, so it remained 1269 two-legged ties. The estimated Nadaraya-Watson estimator ($h_0 =0.539$ by AIC criterion, standard Gaussian kernel) is shown in Figure \ref{fig:phatET}. Compared to the estimator using all data, the two curves are remarkably close around 0. This `no-extra-time' version gives $\hat{p}_{h_0}(0) = 0.540$. So, somewhat surprisingly, the magnitude of the SLHA is actually not influenced at all by the `extra-time' element. What the plot reveals, though, is that the two curves move apart as $x$ moves away from 0. For $x>0$, the dashed line (`no-extra-time' curve) is above the solid line, and conversely for $x<0$. This can be interpreted the following way: when an underdog playing away on the second leg still manages to qualify, that happens `often' after extra-time, i.e., after a long and though battle. Indeed, one doesn't expect an underdog to go and qualify easily at the home ground of a much stronger team.

\subsection{Discussion} \label{subsec:disc}

This analysis investigated the existence of a second leg home advantage in two-stage knockout matches in the UEFA Champions and Europa leagues from 2009/10 to 2014/15. {\it A significant effect was found}. This finding contrasts with other research where the difference in team strength was controlled for, most relevantly \cite{Page2007a}, who found a significant positive SLHA that had disappeared by 1995/96, and \cite{Eugster2011}, who found no effect. Obviously, these conflicting conclusions may have many different origins. First and foremost, the analysed data were not exactly the same. \cite{Page2007a} analysed historical data from all the European cups from their early time (1955/56) up until 2005/06 (taking into account a potential effect of time), while \cite{Eugster2011} looked only at the final knockout stage of the UEFA Champions League from 1994/95 to 2009/10.

\ppn In this paper it was decided to analyse data from 2009/10 to 2014/15 in order to keep the control variable $X$, based on the UEFA club coefficients, homogeneous across the study. Indeed, the way that the UEFA calculates those coefficients have drastically changed over time, most recently in 1999, 2004 and 2009. \cite{Page2007a} acknowledged this issue and addressed it by using indicator variables to create three covariates to include in their logistic model:
\[ \logit(p(x)) = \alpha + \beta_1 \mathbf{I}_{(60/61-98/99)}x + \beta_2\mathbf{I}_{(99/00-03/04)}x + \beta_3\mathbf{I}_{(04/05-05/06)}x \]
(where in their case, $x$ is the difference in UEFA coefficients, as mentioned in Section \ref{subsec:dataprep}). This allowed them to estimate the second-leg home advantage across all seasons as a whole whilst controlling for team strength via the three calculation methods simultaneously. However the usefulness of an overall measure is not particularly high as the strength of the effect is known to be changing within the same time period. Their refined time series analysis of the effect essentially breaks the data down into sub-time periods, anyway. \cite{Eugster2011}, on the other hand, did not account for the differences in coefficient calculation methods, or at least they did not mention anything of that sort. In fact, they incorrectly stated that their entire dataset utilises 20\% of the country coefficient and then assume that there are no changes in calculation method [their Section 2.3]. However, there were three changes in the weight that the country index carries over the period that they studied (see Footnote \ref{foot:coutnry}.). The control variable was also different, as here the log-difference between the UEFA coefficients was used, as opposed to their difference in \cite{Page2007a} and \cite{Eugster2011}. This is arguably more natural for comparing such positive indices. 

\ppn Finally, \cite{Page2007a} and \cite{Eugster2011} both used logistic regression modelling, but neither provided, or made reference to, any diagnostics investigating the validity of their models. Yet, there are serious concerns about it. For instance, Figure 2 in \cite{Eugster2011} indicates that, for a {\it maximum} (normalised) difference in UEFA coefficients ($-1$ or $1$), so in the most unbalanced tie one can imagine, the `underdog' team keeps around 20\% chance of qualifying. Common sense alone suggests that, if La Fiorita was ever to face FC Barcelona, they would not have 1 in 5 chances of qualifying. 

\ppn Actually, testing the goodness-of-fit of the logistic model for the data analysed in this paper lead to reject it. Worse, forcing a logistic-based analysis did not allow a significant SLHA effect to be highlighted, whereas it was picked up by the nonparametric procedure. 
Highlighting a significant SLHA effect opens the door for further analyses. For instance, \cite{Lidor2011} analysed 199 two-legged knockout ties from the Champions and Europa leagues from 1994/95 to 2006/07, and investigated whether the number of goals scored in each leg is associated with the second-leg home advantage. Table \ref{tab:goals_scored_lidor} shows that the average goals scored by the away team in each leg remains similar but the goals scored by the home team increases approximately 33\% from the FLHT to the SLHT. This suggests that the cause of any potential second-leg home advantage is not due to any pressures or influences on the FLHT when playing away on the second leg, but rather it is a boost to the performance of the SLHT on their own soil relative to the FLHT on theirs. Admittedly this analysis did not control for the relative team strengths (see Section \ref{sec:intro}) but it does provide an interesting avenue for future research to explore. Reproducing this type of analysis whilst controlling for team strengths would probably provide a better understanding of whether the effect is an advantage to the SLHT or more so a disadvantage to the FLHT - or both.

\begin{table}[!htbp] 
\centering
\begin{tabular}{@{\extracolsep{5pt}}ccc} 
\hline \\[-1.8ex]
 & First leg & Second leg \\\hline
\\[-1.8ex]
Home Team & 1.27 & 1.68 \\
Away Team & 0.89 & 0.85\\\hline
\end{tabular}
\caption{Average number of goals score by each team.}
\label{tab:goals_scored_lidor}
\end{table}

\ppn More specific questions may be asked as well. For instance, \cite{Carron2005} stated in their review that the home field advantage is apparently universal across all types of sport, yet not universal across all teams {\it within} a sport. Investigating if the second-leg home advantage affects different teams individually would also be of interest. Historically, some clubs have indeed demonstrated expertise in improbable comebacks when playing home on the 
second leg, what is now known in the football folklore as {\it remontada} (Spanish for `comeback' or `catch-up'). Research could also turn to the realm of psychology and sociology, attempting to develop for the SLHA a conceptual framework similar to that of \cite{Courneya1992} and \cite{Carron2005} and briefly exposed in Section \ref{sec:intro}.

\section{Conclusion} \label{sec:chap5}

Motivated by a formal analysis of the existence (or otherwise) of the so-called `{\it second-leg home advantage}' in some international football competitions, this paper aimed to develop better tools for drawing reliable conclusions from a binary regression model, that is, when a conditional probability function $p(x)$ is to be empirically estimated. In particular, a reliable method for constructing pointwise (i.e., for a fixed value of $x$) confidence intervals with good empirical coverage properties was needed. 

\ppn Avoiding rigid and sometimes unwarranted parametric specifications for the function $p$, the method developed here is based on the Nadaraya-Watson estimator, arguably one of the simplest nonparametric regression estimators. In the case of a binary response, this estimator returns a kind of `conditional sample proportion', from which standard confidence intervals of Wald type can easily be constructed. However, in the basic case of estimating a binomial probability, the Wald confidence interval is known to perform very poorly, and alternative confidence intervals, such as the Wilson and the Agresti-Coull intervals, have been strongly recommended.

\ppn The first main methodological contribution of the paper was to extend those `better' confidence intervals to the conditional case. Given that the Nadaraya-Watson estimator is a locally weighted average of the observed binary responses, that extension was very natural and did not present any problem. `Conditional versions' of the Wilson and Agresti-Coull intervals were thus proposed. Actually, any estimator of type 
\begin{equation} \hat{p}_\theta(x) = \sum_{i=1}^n W_i(x;\theta) Y_i, \label{eqn:linest} \end{equation}
where $\{W_i(\cdot;\theta); i=1,\ldots,n\}$ is a set of weights, possibly depending on a parameter $\theta$ (often: a smoothing parameter, such as a bandwidth) and summing to 1, can be regarded as a `local sample proportion', and hence could serve as the basis of the methodology {\it mutatis mutandis}. Nonparametric regression estimators of type (\ref{eqn:linest}) are known as {\it linear smoothers}, and include many common nonparametric regression estimators such as Local Linear, Splines or basic (i.e., without thresholding) Wavelet estimators, for instance. The methodology developed here is thus very general.

\ppn As often when nonparametric function estimation is involved, the inherent bias of estimators like (\ref{eqn:linest}) constitutes a major stumbling block when devising inferential tools. When building confidence intervals, it has been advocated that proceeding via {\it undersmoothing}, that is, working purposely with a sub-optimal bandwidth, would be beneficial in theory. However, an attractive and effective data-driven procedure for selecting such an `undersmoothed' bandwidth was missing up until now. The second main methodological contribution of the paper was to suggest such a procedure, based on some bootstrap resampling scheme. 

\ppn Somewhat surprisingly, the bandwidth returned by the procedure and supposed to be optimal for building good confidence intervals, was not seen to be necessarily smaller than what it should be otherwise. That is in contrast with what a naive interpretation of `undersmoothing' would suggest. The procedure was validated through a simulation study, and proved very efficient at returning a bandwidth guaranteeing empirical coverage very close to the nominal level for the so-constructed intervals in all situations. Importantly, nothing in the procedure pertains to the binary regression framework, so it is clear that the suggested methodology can be used for selecting the right bandwidth for building confidence intervals for a general regression function as well.

\ppn These new intervals were finally used for answering the research question as to the existence of the second-leg home advantage in international football competitions. To that purpose, data from the UEFA Champions and Europa leagues from 2009/10 to 2014/15 were collected and analysed. Working within the regression framework allowed for the abilities of the teams involved to be taken into account, which, due to UEFA seeding regulations, confounds the relationship between playing at home in the second game and the probability of a qualification. This confounding factor was confirmed by an explanatory analysis of the data. For reasons made clear in the paper, the relative strength of the matched teams was measured through the log-difference of the UEFA coefficients of the clubs. Then, the nonparametric model revealed a significant second-leg home advantage, with an estimated probability of qualifying when playing at home on the second leg of $0.539$ and 95\% confidence interval $[0.504;0.574]$, after controlling for the teams' abilities. The existence of such an unwarranted advantage for the team playing at home second may call for some system of compensation and/or handicap in knockout stages of UEFA administered competitions.

\ppn Importantly, the analysis provided is this paper is very objective, in the sense that, purely nonparametric in nature, it does not rely on any arbitrary assumption enforced by the analyst and which could orientate the conclusions in one or the other direction. In particular, no second-leg home advantage effect was evidenced by previous research, exclusively based on parametric models such as logistic regression but without any justification or validation of that parametric specification. It is revealing to observe that, although not fully supported by the data here, a similar analysis based on logistic modelling was not able to highlight the effect. Model misspecification can, indeed, hide interesting features.

\section*{Acknowledgements} 
This research was supported by a Faculty Research Grant from the Faculty of Science, UNSW Sydney (Australia).


\begin{thebibliography}{99}
\bibitem[Agresti and Coull(1998)]{Agresti1998} Agresti, A.\ and Coull, B.A.\ (1998), Approximate is better than ``exact'' for interval estimation of binomial proportions, Amer.\ Statist., 52, 119-126.
\bibitem[Blyth and Still(1983)]{Blyth1983} Blyth, C.R.\ and Still, H.A.\ (1983), Binomial confidence intervals, J.\ Amer.\ Statist.\ Assoc., 78, 108-116.
\bibitem[Brown et al(2001)]{Brown2001b} Brown, L.D., Cai, T.T.\ and DasGupta, A.\ (2001), Interval estimation for a binomial proportion, Statist.\ Sci., 16, 101-133.
\bibitem[Brown et al(2002)]{Brown2002} Brown, L.D., Cai, T.T.\ and DasGupta, A.\ (2002), Confidence intervals for a binomial proportion and asymptotic expansions, Ann.\ Statist., 30, 160-201.
\bibitem[Carron et al(2005)]{Carron2005} Carron, A.V., Loughhead, T.M.\ and Bray, S.R.\ (2005), The home advantage in sports competitions: Courneya and Carron's (1992) conceptual framework a decade later, J.\ Sports Sci., 23, 395-407.
\bibitem[Chen and Qin(2002)]{Chen2002} Chen, S.X.\ and Qin, Y.S.\ (2002), Confidence intervals based on local linear smoother, Scand.\ J.\ Statist., 29, 89-99.
\bibitem[Copas(1983)]{Copas83} Copas, J.B.\ (1983), Plotting $p$ against $x$, J.\ Roy.\ Statist.\ Soc.\ Ser.\ C, 32, 25-31.
\bibitem[Courneya and Carron(1992)]{Courneya1992} Courneya, K.S.\ and Carron, A.V.\ (1992), The home advantage in sport competitions: A literature review, Journal of Sport and Exercise Psychology, 14, 13-27.
\bibitem[Cressie(1978)]{Cressie1978} Cressie, N.\ (1978), A finely tuned continuity correction, Ann.\ Inst.\ Statist.\ Math., 30, 435-442.
\bibitem[Eguchi et al(2003)]{Eguchi03} Eguchi, S., Kim, T.Y.\ and Park, B.U.\ (2003), Local likelihood method: A bridge over parametric and nonparametric regression, J.\ Nonparametr.\ Stat., 15, 665-683.
\bibitem[Eubank and Speckman(1993)]{Eubank1993} Eubank, R.L.\ and Speckman, P.L.\ (1993), Confidence Bands in Nonparametric Regression, J.\ Amer.\ Statist.\ Assoc., 88, 1287-1301.
\bibitem[Eugster et al(2011)]{Eugster2011} Eugster, M.J.A., Gertheiss, J.\ and Kaiser, S.\ (2011), Having the second leg at home: advantage in the UEFA Champions League knockout phase?, J.\ Quant.\ Anal.\ Sports, 7, 1.
\bibitem[Flores et al(2015)]{Flores2015} Flores, R., Forrest, D., de Pablo, C.\ and Tena, J.\ (2015), What is a good result in the first leg of a two-legged football match? European J.\ Oper.\ Res., 247, 641-647.
\bibitem[Ghosh(1979)]{Ghosh1979} Ghosh, B.K.\ (1979), A comparison of some approximate confidence intervals for the binomial parameter, J.\ Amer.\ Statist.\ Assoc., 74, 894-900.
\bibitem[Hall(1992)]{Hall1992} Hall, P.\ (1992), On bootstrap confidence intevals in nonparametric regression, Ann.\ Statist., 20, 695-711.
\bibitem[Hall and Horowitz(2013)]{Hall13} Hall, P.\ and Horowitz, J.\ (2013), A simple bootstrap method for constructing nonparametric confidence bands for functions, Ann.\ Statist., 41, 1892-1921.
\bibitem[H\"ardle and Bowman(1988)]{Hardle1988} H\"ardle, W.K.\ and Bowman, A.W.\ (1988), Bootstrapping in nonparametric regression: local adaptive smoothing and confidence bands, J.\ Amer.\ Statist.\ Assoc., 83, 102-110.
\bibitem[H\"ardle et al(2004)]{Hardle2004} H\"ardle, W.K., M\"uller, M., Sperlich, S.\ and Werwatz, A., Nonparametric and Semiparametric Models: an Introduction, Springer, 2004.
\bibitem[Horowitz and Savin(2001)]{Horowitz2001} Horowitz, J.L. and Savin, N.E. (2001), Binary response models: logits, probits and semiparametrics, J.\ Econ.\ Persp., 15, 43-56.
\bibitem[Hurvich et al(1998)]{Hurvich98} Hurvich, C.M., Simonoff, J.S.\ and Tsai, C.-L.\ (1998), Smoothing parameter selection in nonparametric regression using an improved Akaike information criterion, J.\ R.\ Stat.\ Soc.\ Ser.\ B Stat.\ Methodol., 60, 271-293. 
\bibitem[Jamieson(2010)]{Jamieson2010} Jamieson, J.P.\ (2010), The home field advantage in athletics: a meta-analysis, J.\ Appl.\ Soc.\ Psychol., 40, 1819-1848.
\bibitem[Kassies(2016)]{Kassies} Kassies, B.\ (2016), UEFA European Cup Coefficients Database, {\tt http://kassiesa.home.xs4all.nl/bert/uefa/data/index.html}.
\bibitem[K\"ohler et al(2014)]{Kohler14} K\"ohler, M., Schindler, A.\ and Sperlich, S.\ (2014), A Review and Comparison of Bandwidth Selection Methods for Kernel Regression, Int.\ Stat.\ Rev., 82, 243-274.
\bibitem[le Cessie and van Houwelingen(1991)]{Cessie91} le Cessie, S.\ and van Houwelingen, J.C.\ (1991), A goodness-of-fit test for binary regression models based on smoothing methods, Biometrics, 47, 1267-1282.
\bibitem[Lidor et al(2010)]{Lidor2011} Lidor, R., BarEli, M., Arnon, M.\ and BarEli, A.A.\ (2010), On the advantage of playing the second game at home in the knockout stages of European soccer cup competitions, International Journal of Sport and Exercise Psychology, 8, 312-325.
\bibitem[Nadaraya(1964)]{Nadaraya1964} Nadaraya, E.A.\ (1964), On estimating regression, Theory Probab.\ Appl., 9, 141-142.
\bibitem[Neumann(1997)]{Neumann1997} Neumann, M.H.\ (1997), Pointwise confidence intervals in nonparametric regression with heteroscedastic error structure, Statistics, 29, 1-36.
\bibitem[Nevill and Holder(1999)]{Nevill99} Nevill, A.M.\ and Holder, R.L.\ (1999), Home advantage in sport: an overview of studies on the advantage of playing at home, Sports Med., 28, 221-236.
\bibitem[Olivier and May(2006)]{Olivier2006} Olivier, J.\ and May, W.L.\ (2006), Weighted confidence interval construction for binomial parameters, Stat.\ Methods Med.\ Res., 15, 37-46.
\bibitem[Page and Page(2007)]{Page2007a} Page, L.\ and Page, K.\ (2007), The second leg home advantage: evidence from European football cup competitions, J.\ Sports Sci., 25, 1547-1556.
\bibitem[Pollard(1986)]{Pollard1986} Pollard, R.\ (1986), Home advantage in soccer: a retrospective analysis, J.\ Sports Sci., 4, 237-248.
\bibitem[Pollard(2006)]{Pollard2006a} Pollard, R.\ (2006), Home advantage in soccer: variations in its magnitude and a literature review of the inter-related factors associated with its existence, J.\ Sport Behav., 29, 169-189.\bibitem[Pollard(2008)]{Pollard2008} Pollard, R.\ (2008), Home advantage in football: a current review of an unsolved puzzle, Open Sports Sci.\ J., 1, 12-14.
\bibitem[Pollard and Pollard(2005)]{Pollard2005a} Pollard, R.\ and Pollard, G.\ (2005), Long-term trends in home advantage in professional team sports in North America and England (1876-2003), J.\ Sports Sci., 23, 337-350.
\bibitem[Rodr\'iguez-Campos and Cao-Abad(1993)]{Rodriguez93} Rodr\'iguez-Campos, M.C.\ and Cao-Abad, R.\ (1993), Nonparametric bootstrap confidence intervals for discrete regression functions, J.\ Econometrics, 58, 207-222.
\bibitem[Schwartz and Barsky(1977)]{Schwartz1977} Schwartz, B.\ and Barsky, S.F.\ (1977), The home advantage, Social Forces, 55, 641-661.
\bibitem[Wasserman(2006)]{Wasserman2006} Wasserman, L., All of Nonparametric Statistics, Springer, 2006.
\bibitem[Watson(1964)]{Watson1964} Watson, G.S.\ (1964), Smooth regression analysis, Sankhya, 26, 359-372.
\bibitem[Wilson(1927)]{Wilson1927} Wilson, E.\ (1927), Probable inference, the law of succession, and statistical inference, J.\ Amer.\ Statist.\ Assoc., 22, 209-212.
\bibitem[Xia(1998)]{Xia98} Xia, Y.\ (1998), Bias-corrected confidence bands in nonparametric regression, J.\ R.\ Stat.\ Soc.\ Ser.\ B Stat.\ Methodol., 60, 797-811.
\end{thebibliography}
\end{document}